\documentclass[10pt,journal,compsoc]{IEEEtran}
\usepackage{url}
\usepackage[utf8]{inputenc}
\usepackage{xcolor}
\usepackage{amsmath}
\usepackage{amssymb}
\usepackage{xspace}
\usepackage{epsfig}
\usepackage{balance}

\usepackage[acronyms,nonumberlist,nopostdot,nomain,nogroupskip,acronymlists={hidden}]{glossaries}
\newglossary[algh]{hidden}{acrh}{acnh}{Hidden Acronyms}

\usepackage{booktabs}
\usepackage{tabularx}

\usepackage{tikz}
\usepackage{pgfplots}
\pgfplotsset{compat=newest}
\pgfplotsset{plot coordinates/math parser=false}
\newlength\fheight
\newlength\fwidth
\usetikzlibrary{plotmarks,patterns,decorations.pathreplacing,backgrounds,calc,arrows,arrows.meta,spy,matrix,scopes}
\usepgfplotslibrary{patchplots,groupplots}
\usepackage{tikzscale}
\usepackage[draft]{hyperref}

\newif\ifexttikz
\exttikzfalse

\ifexttikz
	\usetikzlibrary{external}
	\usepackage{fontspec}
\fi

\usepackage{multirow}

\usepackage[font=footnotesize]{subcaption}
\usepackage[font=footnotesize]{caption}

\usepackage{mathtools}
\usepackage[numbers,sort&compress]{natbib}

\usepackage{soul}

\newacronym{3gpp}{3GPP}{3rd Generation Partnership Project}
\newacronym{4g}{4G}{4th generation}
\newacronym{5g}{5G}{5th generation}
\newacronym{6g}{6G}{6th generation}
\newacronym{5gc}{5GC}{5G Core}
\newacronym{adc}{ADC}{Analog to Digital Converter}
\newacronym{aerpaw}{AERPAW}{Aerial Experimentation and Research Platform for Advanced Wireless}
\newacronym{ai}{AI}{Artificial Intelligence}
\newacronym{aimd}{AIMD}{Additive Increase Multiplicative Decrease}
\newacronym{am}{AM}{Acknowledged Mode}
\newacronym{amc}{AMC}{Adaptive Modulation and Coding}
\newacronym{amf}{AMF}{Access and Mobility Management Function}
\newacronym{aops}{AOPS}{Adaptive Order Prediction Scheduling}
\newacronym{api}{API}{Application Programming Interface}
\newacronym{apn}{APN}{Access Point Name}
\newacronym{ap}{AP}{Application Protocol}
\newacronym{aqm}{AQM}{Active Queue Management}
\newacronym{ausf}{AUSF}{Authentication Server Function}
\newacronym{avc}{AVC}{Advanced Video Coding}
\newacronym{awgn}{AGWN}{Additive White Gaussian Noise}
\newacronym{balia}{BALIA}{Balanced Link Adaptation Algorithm}
\newacronym{bbu}{BBU}{Base Band Unit}
\newacronym{bdp}{BDP}{Bandwidth-Delay Product}
\newacronym{ber}{BER}{Bit Error Rate}
\newacronym{bf}{BF}{Beamforming}
\newacronym{bler}{BLER}{Block Error Rate}
\newacronym{brr}{BRR}{Bayesian Ridge Regressor}
\newacronym{bs}{BS}{Base Station}
\newacronym{bsr}{BSR}{Buffer Status Report}
\newacronym{bss}{BSS}{Business Support System}
\newacronym{ca}{CA}{Carrier Aggregation}
\newacronym{caas}{CaaS}{Connectivity-as-a-Service}
\newacronym{cb}{CB}{Code Block}
\newacronym{cc}{CC}{Congestion Control}
\newacronym{ccid}{CCID}{Congestion Control ID}
\newacronym{cco}{CC}{Carrier Component}
\newacronym{cdd}{CDD}{Cyclic Delay Diversity}
\newacronym{cdf}{CDF}{Cumulative Distribution Function}
\newacronym{cdn}{CDN}{Content Distribution Network}
\newacronym{cn}{CN}{Core Network}
\newacronym{codel}{CoDel}{Controlled Delay Management}
\newacronym{comac}{COMAC}{Converged Multi-Access and Core}
\newacronym{cord}{CORD}{Central Office Re-architected as a Datacenter}
\newacronym{cornet}{CORNET}{COgnitive Radio NETwork}
\newacronym{cosmos}{COSMOS}{Cloud Enhanced Open Software Defined Mobile Wireless Testbed for City-Scale Deployment}
\newacronym{cots}{COTS}{Commercial Off-the-Shelf}
\newacronym{cp}{CP}{Control Plane}
\newacronym{cpu}{CPU}{Central Processing Unit}
\newacronym{cqi}{CQI}{Channel Quality Information}
\newacronym{cr}{CR}{Cognitive Radio}
\newacronym{cran}{CRAN}{Cloud \gls{ran}}
\newacronym{crs}{CRS}{Cell Reference Signal}
\newacronym{csi}{CSI}{Channel State Information}
\newacronym{csirs}{CSI-RS}{Channel State Information - Reference Signal}
\newacronym{cu}{CU}{Central Unit}
\newacronym{d2tcp}{D$^2$TCP}{Deadline-aware Data center TCP}
\newacronym{d3}{D$^3$}{Deadline-Driven Delivery}
\newacronym{dac}{DAC}{Digital to Analog Converter}
\newacronym{dag}{DAG}{Directed Acyclic Graph}
\newacronym{das}{DAS}{Distributed Antenna System}
\newacronym{dash}{DASH}{Dynamic Adaptive Streaming over HTTP}
\newacronym{dc}{DC}{Dual Connectivity}
\newacronym{dccp}{DCCP}{Datagram Congestion Control Protocol}
\newacronym{dce}{DCE}{Direct Code Execution}
\newacronym{dci}{DCI}{Downlink Control Information}
\newacronym{dctcp}{DCTCP}{Data Center TCP}
\newacronym{dl}{DL}{Downlink}
\newacronym{dmr}{DMR}{Deadline Miss Ratio}
\newacronym{dmrs}{DMRS}{DeModulation Reference Signal}
\newacronym{drlcc}{DRL-CC}{Deep Reinforcement Learning Congestion Control}
\newacronym{drs}{DRS}{Discovery Reference Signal}
\newacronym{du}{DU}{Distributed Unit}
\newacronym{e2e}{E2E}{end-to-end}
\newacronym{ecaas}{ECaaS}{Edge-Cloud-as-a-Service}
\newacronym{ecn}{ECN}{Explicit Congestion Notification}
\newacronym{edf}{EDF}{Earliest Deadline First}
\newacronym{embb}{eMBB}{Enhanced Mobile Broadband}
\newacronym{empower}{EMPOWER}{EMpowering transatlantic PlatfOrms for advanced WirEless Research}
\newacronym{enb}{eNB}{evolved Node Base}
\newacronym{endc}{EN-DC}{E-UTRAN-\gls{nr} \gls{dc}}
\newacronym{epc}{EPC}{Evolved Packet Core}
\newacronym{eps}{EPS}{Evolved Packet System}
\newacronym{es}{ES}{Edge Server}
\newacronym{etsi}{ETSI}{European Telecommunications Standards Institute}
\newacronym[firstplural=Estimated Times of Arrival (ETAs)]{eta}{ETA}{Estimated Time of Arrival}
\newacronym{eutran}{E-UTRAN}{Evolved Universal Terrestrial Access Network}
\newacronym{faas}{FaaS}{Function-as-a-Service}
\newacronym{fapi}{FAPI}{Functional Application Platform Interface}
\newacronym{fdd}{FDD}{Frequency Division Duplexing}
\newacronym{fdm}{FDM}{Frequency Division Multiplexing}
\newacronym{fdma}{FDMA}{Frequency Division Multiple Access}
\newacronym{fed4fire}{FED4FIRE+}{Federation 4 Future Internet Research and Experimentation Plus}
\newacronym{fir}{FIR}{Finite Impulse Response}
\newacronym{fit}{FIT}{Future \acrlong{iot}}
\newacronym{fpga}{FPGA}{Field Programmable Gate Array}
\newacronym{fr2}{FR2}{Frequency Range 2}
\newacronym{fs}{FS}{Fast Switching}
\newacronym{fscc}{FSCC}{Flow Sharing Congestion Control}
\newacronym{ftp}{FTP}{File Transfer Protocol}
\newacronym{fw}{FW}{Flow Window}
\newacronym{ge}{GE}{Gaussian Elimination}
\newacronym{gnb}{gNB}{Next Generation Node Base}
\newacronym{gop}{GOP}{Group of Pictures}
\newacronym{gpr}{GPR}{Gaussian Process Regressor}
\newacronym{gpu}{GPU}{Graphics Processing Unit}
\newacronym{gtp}{GTP}{GPRS Tunneling Protocol}
\newacronym{gtpc}{GTP-C}{GPRS Tunnelling Protocol Control Plane}
\newacronym{gtpu}{GTP-U}{GPRS Tunnelling Protocol User Plane}
\newacronym{gtpv2c}{GTPv2-C}{\gls{gtp} v2 - Control}
\newacronym{gw}{GW}{Gateway}
\newacronym{harq}{HARQ}{Hybrid Automatic Repeat reQuest}
\newacronym{hetnet}{HetNet}{Heterogeneous Network}
\newacronym{hh}{HH}{Hard Handover}
\newacronym{hol}{HOL}{Head-of-Line}
\newacronym{hqf}{HQF}{Highest-quality-first}
\newacronym{hss}{HSS}{Home Subscription Server}
\newacronym{http}{HTTP}{HyperText Transfer Protocol}
\newacronym{ia}{IA}{Initial Access}
\newacronym{iab}{IAB}{Integrated Access and Backhaul}
\newacronym{ic}{IC}{Incident Command}
\newacronym{ietf}{IETF}{Internet Engineering Task Force}
\newacronym{imsi}{IMSI}{International Mobile Subscriber Identity}
\newacronym{imt}{IMT}{International Mobile Telecommunication}
\newacronym{iot}{IoT}{Internet of Things}
\newacronym{ip}{IP}{Internet Protocol}
\newacronym{itu}{ITU}{International Telecommunication Union}
\newacronym{kpi}{KPI}{Key Performance Indicator}
\newacronym{kpm}{KPM}{Key Performance Measurement}
\newacronym{kvm}{KVM}{Kernel-based Virtual Machine}
\newacronym{los}{LOS}{Line-of-Sight}
\newacronym{lsm}{LSM}{Link-to-System Mapping}
\newacronym{lstm}{LSTM}{Long Short Term Memory}
\newacronym{lte}{LTE}{Long Term Evolution}
\newacronym{lxc}{LXC}{Linux Container}
\newacronym{m2m}{M2M}{Machine to Machine}
\newacronym{mac}{MAC}{Medium Access Control}
\newacronym{manet}{MANET}{Mobile Ad Hoc Network}
\newacronym{mano}{MANO}{Management and Orchestration}
\newacronym{mc}{MC}{Multi-Connectivity}
\newacronym{mcc}{MCC}{Mobile Cloud Computing}
\newacronym{mchem}{MCHEM}{Massive Channel Emulator}
\newacronym{mcs}{MCS}{Modulation and Coding Scheme}
\newacronym{mec}{MEC}{Multi-access Edge Computing}
\newacronym{mec2}{MEC}{Mobile Edge Cloud}
\newacronym{mfc}{MFC}{Mobile Fog Computing}
\newacronym{mgen}{MGEN}{Multi-Generator}
\newacronym{mi}{MI}{Mutual Information}
\newacronym{mib}{MIB}{Master Information Block}
\newacronym{miesm}{MIESM}{Mutual Information Based Effective SINR}
\newacronym{mimo}{MIMO}{Multiple Input, Multiple Output}
\newacronym{ml}{ML}{Machine Learning}
\newacronym{mlr}{MLR}{Maximum-local-rate}
\newacronym[plural=\gls{mme}s,firstplural=Mobility Management Entities (MMEs)]{mme}{MME}{Mobility Management Entity}
\newacronym{mmtc}{mMTC}{Massive Machine-Type Communications}
\newacronym{mmwave}{mmWave}{millimeter wave}
\newacronym{mpdccp}{MP-DCCP}{Multipath Datagram Congestion Control Protocol}
\newacronym{mptcp}{MPTCP}{Multipath TCP}
\newacronym{mr}{MR}{Maximum Rate}
\newacronym{mrdc}{MR-DC}{Multi \gls{rat} \gls{dc}}
\newacronym{mse}{MSE}{Mean Square Error}
\newacronym{mss}{MSS}{Maximum Segment Size}
\newacronym{mt}{MT}{Mobile Termination}
\newacronym{mtd}{MTD}{Machine-Type Device}
\newacronym{mtu}{MTU}{Maximum Transmission Unit}
\newacronym{mumimo}{MU-MIMO}{Multi-user \gls{mimo}}
\newacronym{mvno}{MVNO}{Mobile Virtual Network Operator}
\newacronym{nalu}{NALU}{Network Abstraction Layer Unit}
\newacronym{nas}{NAS}{Non-Access Stratum}
\newacronym{nbiot}{NB-IoT}{Narrow Band IoT}
\newacronym{nfv}{NFV}{Network Function Virtualization}
\newacronym{nfvi}{NFVI}{Network Function Virtualization Infrastructure}
\newacronym{nic}{NIC}{Network Interface Card}
\newacronym{nlos}{NLOS}{Non-Line-of-Sight}
\newacronym{now}{NOW}{Non Overlapping Window}
\newacronym{nsm}{NSM}{Network Service Mesh}
\newacronym[type=hidden]{nr}{NR}{New Radio}
\newacronym{nrf}{NRF}{Network Repository Function}
\newacronym{nsa}{NSA}{Non Stand Alone}
\newacronym{nse}{NSE}{Network Slicing Engine}
\newacronym{nssf}{NSSF}{Network Slice Selection Function}
\newacronym{o2i}{O2I}{Outdoor to Indoor}
\newacronym{oai}{OAI}{OpenAirInterface}
\newacronym{oaicn}{OAI-CN}{\gls{oai} \acrlong{cn}}
\newacronym{oairan}{OAI-RAN}{\acrlong{oai} \acrlong{ran}}
\newacronym{oam}{OAM}{Operations, Administration and Maintenance}
\newacronym{ofdm}{OFDM}{Orthogonal Frequency Division Multiplexing}
\newacronym{olia}{OLIA}{Opportunistic Linked Increase Algorithm}
\newacronym{omec}{OMEC}{Open Mobile Evolved Core}
\newacronym{onap}{ONAP}{Open Network Automation Platform}
\newacronym{onf}{ONF}{Open Networking Foundation}
\newacronym{onos}{ONOS}{Open Networking Operating System}
\newacronym{oom}{OOM}{\gls{onap} Operations Manager}
\newacronym{opnfv}{OPNFV}{Open Platform for \gls{nfv}}
\newacronym[type=hidden]{oran}{O-RAN}{Open \gls{ran}}
\newacronym{orbit}{ORBIT}{Open-Access Research Testbed for Next-Generation Wireless Networks}
\newacronym{os}{OS}{Operating System}
\newacronym{oss}{OSS}{Operations Support System}
\newacronym{pa}{PA}{Position-aware}
\newacronym{pase}{PASE}{Prioritization, Arbitration, and Self-adjusting Endpoints}
\newacronym{pawr}{PAWR}{Platforms for Advanced Wireless Research}
\newacronym{pbch}{PBCH}{Physical Broadcast Channel}
\newacronym{pcef}{PCEF}{Policy and Charging Enforcement Function}
\newacronym{pcfich}{PCFICH}{Physical Control Format Indicator Channel}
\newacronym{pcrf}{PCRF}{Policy and Charging Rules Function}
\newacronym{pdcch}{PDCCH}{Physical Downlink Control Channel}
\newacronym{pdcp}{PDCP}{Packet Data Convergence Protocol}
\newacronym{pdsch}{PDSCH}{Physical Downlink Shared Channel}
\newacronym{pdu}{PDU}{Packet Data Unit}
\newacronym{pf}{PF}{Proportional Fair}
\newacronym{pgw}{PGW}{Packet Gateway}
\newacronym{phich}{PHICH}{Physical Hybrid ARQ Indicator Channel}
\newacronym{phy}{PHY}{Physical}
\newacronym{pmch}{PMCH}{Physical Multicast Channel}
\newacronym{pmi}{PMI}{Precoding Matrix Indicators}
\newacronym{powder}{POWDER}{Platform for Open Wireless Data-driven Experimental Research}
\newacronym{ppo}{PPO}{Proximal Policy Optimization}
\newacronym{ppp}{PPP}{Poisson Point Process}
\newacronym{prach}{PRACH}{Physical Random Access Channel}
\newacronym{prb}{PRB}{Physical Resource Block}
\newacronym{psnr}{PSNR}{Peak Signal to Noise Ratio}
\newacronym{pss}{PSS}{Primary Synchronization Signal}
\newacronym{pucch}{PUCCH}{Physical Uplink Control Channel}
\newacronym{pusch}{PUSCH}{Physical Uplink Shared Channel}
\newacronym{qam}{QAM}{Quadrature Amplitude Modulation}
\newacronym{qci}{QCI}{\gls{qos} Class Identifier}
\newacronym{qoe}{QoE}{Quality of Experience}
\newacronym{qos}{QoS}{Quality of Service}
\newacronym{quic}{QUIC}{Quick UDP Internet Connections}
\newacronym{rach}{RACH}{Random Access Channel}
\newacronym{ran}{RAN}{Radio Access Network}
\newacronym[firstplural=Radio Access Technologies (RATs)]{rat}{RAT}{Radio Access Technology}
\newacronym{rcn}{RCN}{Research Coordination Network}
\newacronym{rec}{REC}{Radio Edge Cloud}
\newacronym{red}{RED}{Random Early Detection}
\newacronym{renew}{RENEW}{Reconfigurable Eco-system for Next-generation End-to-end Wireless}
\newacronym{rf}{RF}{Radio Frequency}
\newacronym{rfc}{RFC}{Request for Comments}
\newacronym{rfr}{RFR}{Random Forest Regressor}
\newacronym{ric}{RIC}{\gls{ran} Intelligent Controller}
\newacronym{rlc}{RLC}{Radio Link Control}
\newacronym{rlf}{RLF}{Radio Link Failure}
\newacronym{rlnc}{RLNC}{Random Linear Network Coding}
\newacronym{rmr}{RMR}{RIC Message Router}
\newacronym{rmse}{RMSE}{Root Mean Squared Error}
\newacronym{rnis}{RNIS}{Radio Network Information Service}
\newacronym{rr}{RR}{Round Robin}
\newacronym{rrc}{RRC}{Radio Resource Control}
\newacronym{rrm}{RRM}{Radio Resource Management}
\newacronym{rru}{RRU}{Remote Radio Unit}
\newacronym{rs}{RS}{Remote Server}
\newacronym{rsrp}{RSRP}{Reference Signal Received Power}
\newacronym{rsrq}{RSRQ}{Reference Signal Received Quality}
\newacronym{rss}{RSS}{Received Signal Strength}
\newacronym{rssi}{RSSI}{Received Signal Strength Indicator}
\newacronym{rtt}{RTT}{Round Trip Time}
\newacronym{ru}{RU}{Radio Unit}
\newacronym{rw}{RW}{Receive Window}
\newacronym{rx}{RX}{Receiver}
\newacronym{s1ap}{S1AP}{S1 Application Protocol}
\newacronym{sa}{SA}{standalone}
\newacronym{sack}{SACK}{Selective Acknowledgment}
\newacronym{sap}{SAP}{Service Access Point}
\newacronym{sc2}{SC2}{Spectrum Collaboration Challenge}
\newacronym{scef}{SCEF}{Service Capability Exposure Function}
\newacronym{sch}{SCH}{Secondary Cell Handover}
\newacronym{scoot}{SCOOT}{Split Cycle Offset Optimization Technique}
\newacronym{sctp}{SCTP}{Stream Control Transmission Protocol}
\newacronym{sdap}{SDAP}{Service Data Adaptation Protocol}
\newacronym{sdk}{SDK}{Software Development Kit}
\newacronym{sdm}{SDM}{Space Division Multiplexing}
\newacronym{sdma}{SDMA}{Spatial Division Multiple Access}
\newacronym{sdn}{SDN}{Software-defined Networking}
\newacronym{sdr}{SDR}{Software-defined Radio}
\newacronym{seba}{SEBA}{SDN-Enabled Broadband Access}
\newacronym{sgsn}{SGSN}{Serving GPRS Support Node}
\newacronym{sgw}{SGW}{Service Gateway}
\newacronym{si}{SI}{Study Item}
\newacronym{sib}{SIB}{Secondary Information Block}
\newacronym{sinr}{SINR}{Signal to Interference plus Noise Ratio}
\newacronym{sip}{SIP}{Session Initiation Protocol}
\newacronym{siso}{SISO}{Single Input, Single Output}
\newacronym{sla}{SLA}{Service Level Agreement}
\newacronym{sm}{SM}{Service Model}
\newacronym{smf}{SMF}{Session Management Function}
\newacronym{smo}{SMO}{Service Management and Orchestration}
\newacronym{sms}{SMS}{Short Message Service}
\newacronym{smsgmsc}{SMS-GMSC}{\gls{sms}-Gateway}
\newacronym{snr}{SNR}{Signal-to-Noise-Ratio}
\newacronym{son}{SON}{Self-Organizing Network}
\newacronym{sptcp}{SPTCP}{Single Path TCP}
\newacronym{srb}{SRB}{Service Radio Bearer}
\newacronym{srn}{SRN}{Standard Radio Node}
\newacronym{srs}{SRS}{Sounding Reference Signal}
\newacronym{ss}{SS}{Synchronization Signal}
\newacronym{sss}{SSS}{Secondary Synchronization Signal}
\newacronym{st}{ST}{Spanning Tree}
\newacronym{svc}{SVC}{Scalable Video Coding}
\newacronym{tb}{TB}{Transport Block}
\newacronym{tcp}{TCP}{Transmission Control Protocol}
\newacronym{tdd}{TDD}{Time Division Duplexing}
\newacronym{tdm}{TDM}{Time Division Multiplexing}
\newacronym{tdma}{TDMA}{Time Division Multiple Access}
\newacronym{tfl}{TfL}{Transport for London}
\newacronym{tfrc}{TFRC}{TCP-Friendly Rate Control}
\newacronym{tft}{TFT}{Traffic Flow Template}
\newacronym{tgen}{TGEN}{Traffic Generator}
\newacronym{tip}{TIP}{Telecom Infra Project}
\newacronym{tm}{TM}{Transparent Mode}
\newacronym{to}{TO}{Telco Operator}
\newacronym{tr}{TR}{Technical Report}
\newacronym{trp}{TRP}{Transmitter Receiver Pair}
\newacronym{ts}{TS}{Technical Specification}
\newacronym{tti}{TTI}{Transmission Time Interval}
\newacronym{ttt}{TTT}{Time-to-Trigger}
\newacronym{tx}{TX}{Transmitter}
\newacronym{uas}{UAS}{Unmanned Aerial System}
\newacronym{uav}{UAV}{Unmanned Aerial Vehicle}
\newacronym{udm}{UDM}{Unified Data Management}
\newacronym{udp}{UDP}{User Datagram Protocol}
\newacronym{udr}{UDR}{Unified Data Repository}
\newacronym{ue}{UE}{User Equipment}
\newacronym{uhd}{UHD}{\gls{usrp} Hardware Driver}
\newacronym{ul}{UL}{Uplink}
\newacronym{um}{UM}{Unacknowledged Mode}
\newacronym{uml}{UML}{Unified Modeling Language}
\newacronym{upa}{UPA}{Uniform Planar Array}
\newacronym{upf}{UPF}{User Plane Function}
\newacronym{urllc}{URLLC}{Ultra Reliable and Low Latency Communications}
\newacronym{usa}{U.S.}{United States}
\newacronym{usim}{USIM}{Universal Subscriber Identity Module}
\newacronym{usrp}{USRP}{Universal Software Radio Peripheral}
\newacronym{utc}{UTC}{Urban Traffic Control}
\newacronym{vim}{VIM}{Virtualization Infrastructure Manager}
\newacronym{vm}{VM}{Virtual Machine}
\newacronym{vnf}{VNF}{Virtual Network Function}
\newacronym{volte}{VoLTE}{Voice over \gls{lte}}
\newacronym{voltha}{VOLTHA}{Virtual OLT HArdware Abstraction}
\newacronym{vr}{VR}{Virtual Reality}
\newacronym{vran}{vRAN}{Virtualized \gls{ran}}
\newacronym{vss}{VSS}{Video Streaming Server}
\newacronym{wbf}{WBF}{Wired Bias Function}
\newacronym{wf}{WF}{Waterfilling}
\newacronym{wlan}{WLAN}{Wireless Local Area Network}
\newacronym{osm}{OSM}{Open Source \gls{nfv} Management and Orchestration}
\newacronym{pnf}{PNF}{Physical Network Function}
\newacronym{drl}{DRL}{Deep Reinforcement Learning}
\newacronym{mtc}{MTC}{Machine-type Communications}
\newacronym{osc}{OSC}{O-RAN Software Community}
\newacronym{rc}{RC}{RAN Control}
\newacronym{dqn}{DQN}{Deep Q-Network}
\tikzstyle{startstop} = [rectangle, rounded corners, minimum width=2cm, minimum height=0.5cm,text centered, draw=black]
\tikzstyle{io} = [trapezium, trapezium left angle=70, trapezium right angle=110, minimum width=3cm, minimum height=1cm, text centered, draw=black]
\tikzstyle{process} = [rectangle, minimum width=2cm, minimum height=0.5cm, text centered, draw=black, alignb=center]
\tikzstyle{decision} = [ellipse, minimum width=2cm, minimum height=1cm, text centered, draw=black]
\tikzstyle{arrow} = [thick,<->,>=stealth]
\tikzstyle{line} = [thick,>=stealth]
\tikzstyle{darrow} = [thick,<->,>=stealth,dashed]
\tikzstyle{sarrow} = [thick,->,>=stealth]
\tikzstyle{larrow} = [line width=0.1mm,dashdotted,->,>=stealth]
\tikzstyle{llarrow} = [line width=0.1mm,->,>=stealth]

\makeatletter
\def\grd@save@target#1{%
  \def\grd@target{#1}}
\def\grd@save@start#1{%
  \def\grd@start{#1}}
\tikzset{
  grid with coordinates/.style={
    to path={%
      \pgfextra{%
        \edef\grd@@target{(\tikztotarget)}%
        \tikz@scan@one@point\grd@save@target\grd@@target\relax
        \edef\grd@@start{(\tikztostart)}%
        \tikz@scan@one@point\grd@save@start\grd@@start\relax
        \draw[minor help lines] (\tikztostart) grid (\tikztotarget);
        \draw[major help lines] (\tikztostart) grid (\tikztotarget);
        \grd@start
        \pgfmathsetmacro{\grd@xa}{\the\pgf@x/1cm}
        \pgfmathsetmacro{\grd@ya}{\the\pgf@y/1cm}
        \grd@target
        \pgfmathsetmacro{\grd@xb}{\the\pgf@x/1cm}
        \pgfmathsetmacro{\grd@yb}{\the\pgf@y/1cm}
        \pgfmathsetmacro{\grd@xc}{\grd@xa + \pgfkeysvalueof{/tikz/grid with coordinates/major step x}}
        \pgfmathsetmacro{\grd@yc}{\grd@ya + \pgfkeysvalueof{/tikz/grid with coordinates/major step y}}
        \foreach \x in {\grd@xa,\grd@xc,...,\grd@xb}
        \node[anchor=north] at (\x,\grd@ya) {\pgfmathprintnumber{\x}};
        \foreach \y in {\grd@ya,\grd@yc,...,\grd@yb}
        \node[anchor=east] at (\grd@xa,\y) {\pgfmathprintnumber{\y}};
      }
    }
  },
  minor help lines/.style={
    help lines,
    gray,
    line cap =round,
    xstep=\pgfkeysvalueof{/tikz/grid with coordinates/minor step x},
    ystep=\pgfkeysvalueof{/tikz/grid with coordinates/minor step y}
  },
  major help lines/.style={
    help lines,
    line cap =round,
    line width=\pgfkeysvalueof{/tikz/grid with coordinates/major line width},
    xstep=\pgfkeysvalueof{/tikz/grid with coordinates/major step x},
    ystep=\pgfkeysvalueof{/tikz/grid with coordinates/major step y}
  },
  grid with coordinates/.cd,
  minor step x/.initial=.5,
  minor step y/.initial=.2,
  major step x/.initial=1,
  major step y/.initial=1,
  major line width/.initial=1pt,
}
\makeatother

\definecolor{desireRed}{RGB}{230,57,60}%
\definecolor{darkPurple}{RGB}{59,31,43}%
\definecolor{springGreen}{RGB}{37,223,145}%
\definecolor{queenBlue}{RGB}{69,123,157}%
\definecolor{spaceCadet}{RGB}{29,53,87}%

\newcommand{\coloran}{ColO-RAN\xspace}

\usepackage{dblfloatfix}

\begin{document}
\bstctlcite{BSTcontrol}  



\title{\coloran: Developing Machine Learning-based xApps for Open RAN Closed-loop Control on Programmable Experimental Platforms}


\author{\IEEEauthorblockN{Michele Polese,~\IEEEmembership{Member, IEEE}, Leonardo Bonati,~\IEEEmembership{Student Member, IEEE},\\Salvatore D'Oro,~\IEEEmembership{Member, IEEE}, Stefano Basagni,~\IEEEmembership{Senior Member, IEEE},\\Tommaso Melodia,~\IEEEmembership{Fellow, IEEE}}
\thanks{The authors are with the Institute for the Wireless Internet of Things, Northeastern University, Boston, MA, USA. E-mail: \{m.polese, l.bonati, s.doro, s.basagni, t.melodia\}@northeastern.edu.}
\thanks{This work was partially supported by the U.S.\ National Science Foundation under Grants CNS-1923789, CNS-1925601, CNS-2120447, and CNS-2112471, and the U.S.\ Office of Naval Research under Grant N00014-20-1-2132.}
}


\flushbottom
\setlength{\parskip}{0ex plus0.1ex}

\maketitle
\glsunset{nr}
\glsunset{lte}
\glsunset{3gpp}

\begin{abstract}
Cellular networks are undergoing a radical transformation toward disaggregated, fully virtualized, and programmable architectures with increasingly heterogeneous devices and applications. 
In this context, the open architecture standardized by the O-RAN Alliance enables algorithmic and hardware-independent \gls{ran} adaptation through closed-loop control. 
O-RAN introduces \gls{ml}-based network control and automation algorithms as so-called \textit{xApps} running on \acrlongpl{ric}. 
However, in spite of the new opportunities brought about by the Open RAN, advances in \gls{ml}-based network automation have been slow, mainly because of the unavailability of large-scale datasets and experimental testing infrastructure.
This slows down the development and widespread adoption of \gls{drl} agents on real networks, delaying progress in intelligent and autonomous \gls{ran} control.
In this paper, we address these challenges by proposing practical solutions and software pipelines for the design, training, testing, and experimental evaluation of \gls{drl}-based closed-loop control in the Open \gls{ran}.
We introduce \coloran, the first publicly-available large-scale O-RAN testing framework with software-defined radios-in-the-loop.
Building on the scale and computational capabilities of the Colosseum wireless network emulator, \coloran enables \gls{ml} research at scale using O-RAN components, programmable base stations, and a ``wireless data factory''.
%
%
%
Specifically, we design and develop three exemplary xApps for \gls{drl}-based control of RAN slicing, scheduling and online model training, and evaluate their performance on a cellular network with 7 softwarized base stations and 42 users. 
%
%
%
Finally, we showcase the portability of \coloran to different platforms by deploying it on Arena, an indoor programmable testbed.
Extensive results from our first-of-its-kind large-scale evaluation highlight the benefits and challenges of \gls{drl}-based adaptive control. They also provide insights on the development of wireless \gls{drl} pipelines, from data analysis to the design of \gls{drl} agents, and on the tradeoffs associated to training on a live \gls{ran}.
\coloran and the collected large-scale dataset will be made publicly available to the research community.
\end{abstract}


\begin{picture}(0,0)(10,-650)
\put(0,0){
\put(0,0){\footnotesize This work has been submitted to the IEEE for possible publication.}
\put(0,-10){
\footnotesize Copyright may be transferred without notice, after which this version may no longer be accessible.}}
\end{picture}

\glsresetall
\glsunset{nr}
\glsunset{lte}
\glsunset{3gpp}

\section{Introduction}
\label{sec:intro}

In addition to providing traditional voice and data connectivity services, cellular systems are becoming increasingly pervasive in industrial and agricultural automation, interconnecting millions of sensors, vehicles, airplanes, and drones, and providing the nervous system for a plethora of smart systems~\cite{ericsson2021mobility,giordani2020toward}. 
These diverse use cases, however, often come with heterogeneous---possibly orthogonal---network constraints and requirements~\cite{xiong2019deep}. For instance, autonomous driving applications require \gls{urllc} to allow vehicles to promptly react to sudden events and changing traffic conditions. Instead, high-quality multimedia content requires high data rates, but can tolerate a higher packet loss and latency.
Therefore, the future generations of cellular networks need to be \textit{flexible} and \textit{adaptive} to many different application and user requirements.
To achieve these goals, future \glspl{ran} will need to combine three key ingredients~\cite{bonati2020open}: (i)~\textit{programmable and virtualized protocol stacks} with clearly defined, open interfaces; (ii)~\textit{closed-loop network control}, and (iii)~\textit{data-driven modeling and \gls{ml}}.
Programmability will allow swift adaptation of the \gls{ran} to provide bespoke services able to satisfy the requirements of specific deployments.
Closed-loop control will leverage telemetry 
measurements
from the \gls{ran} to reconfigure cellular nodes,
adapting their behavior to 
current
network conditions and traffic.
Last, data-driven modeling will exploit recent developments in \gls{ml} and big data to enable real-time, closed-loop, and dynamic decision-making  based, for instance, on \gls{drl}~\cite{oran-ml}.
These are the very same principles at the core of the Open \gls{ran} paradigm, which has recently gained traction as a practical enabler of algorithmic and hardware innovation in future cellular networks~\cite{niknam2020intelligent,bonati2021intelligence,zhou2021ran}. 
To promote the evolution toward open \gls{ran} architectures, \gls{3gpp} has standardized disaggregated base stations that are split into a number of different functional units, the \gls{cu}, \gls{du}, and \gls{ru}. The O-RAN Alliance, an industry consortium, is standardizing open interfaces that connect the various disaggregated functional units to a common control overlay, the 
\gls{ric}, capable of executing custom control logic via so-called \textit{xApps}.
%
Ultimately, these efforts will render the \textit{monolithic \gls{ran} ``black-box''}
obsolete, favoring
\textit{open, programmable and virtualized} solutions that expose status and offer control knobs through standardized interfaces~\cite{bonati2020open}.

%
Intelligent, dynamic network optimization via add-on software xApps is clearly a key enabler for future network automation. However, it also introduces novel practical challenges concerning, for instance, the deployment of data-driven \gls{ml} control solutions at scale.
%
%
%
Domain-specific challenges stem from considering the constraints of standardized
\glspl{ran}, the very nature of the wireless ecosystem and the complex interplay among  different elements of the networking stack.
These challenges, all yet to be addressed in practical \gls{ran} deployments, include:

%

\noindent \textit{1) Collecting datasets at scale}. Datasets for \gls{ml} training/testing at scale need to be carefully collected and curated to accurately represent the intrinsic randomness and behavior of real-world~\glspl{ran}.

\noindent \textit{2) Testing \gls{ml}-based control at scale}.
Even if \gls{ml} algorithms are trained on properly collected data, 
it is necessary to assess their robustness at scale, especially when considering closed-loop control, to prevent 
poorly-designed
data-driven solutions from causing outages or sub-optimal performance.

\noindent \textit{3) Designing efficient \gls{ml} agents with unreliable input and constrained output}. In production systems, real-time collection of data from the \gls{ran} may be inconsistent (e.g., with varying periodicity) or incomplete (e.g., missing entries), and control actions may be constrained by standard specification.

\noindent \textit{4) Designing \gls{ml} agents capable of generalizing}. Agents should be able to generalize and adapt to unseen deployment configurations not part of the training set.

\noindent \textit{5) Selecting meaningful features}. Features should be accurately selected to provide a meaningful representation of the network status without incurring into dimensionality issues.

%


\textbf{Contributions. }
To address these key challenges,
in this paper we describe the design of \gls{drl}-based xApps for closed-loop control in O-RAN and their testing in a first-of-its-kind softwarized pipeline on large-scale experimental platforms. 
%
%
%
Based on this experience, we review and discuss key insights in the domain of \gls{ml} design for O-RAN networks.
Notably, our contributions are as follows:


\smallskip
\noindent $\bullet$ We introduce \coloran, a first-of-its-kind open, large-scale, experimental O-RAN framework for training and testing \gls{ml} solutions for next-generation \glspl{ran}. It combines O-RAN components, a softwarized
\gls{ran} framework~\cite{bonati2021scope}, and Colosseum, the world's largest, open, and publicly-available wireless network emulator based on \glspl{sdr}~\cite{bonati2021colosseum}. Specifically, \coloran leverages Colosseum as a {\em wireless data factory} to generate large-scale datasets for \gls{ml} training in a variety of \gls{rf} environments, taking into account propagation and fading characteristics of real-world deployments.
The \gls{ml} models are deployed as xApps on the near-real-time \gls{ric}, which connects to \gls{ran} nodes through O-RAN-compliant interfaces for data collection and closed-loop control.
\coloran is the first platform that enables wireless researchers to deploy \gls{ml} solutions on a full-stack, fully virtualized O-RAN environment which integrates large-scale data collection and \gls{drl} testing capabilities with \glspl{sdr}. Moreover, \coloran also offers a lightweight, containerized implementation that can be easily ported to other experimental platforms.
\textit{\coloran and the dataset created for this paper will be open-sourced and made publicly available to the research community.}

\smallskip
\noindent $\bullet$ We develop three xApps for closed-loop control of \gls{ran} scheduling and slicing policies, and for the online training of \gls{drl} agents on live production environments. We propose an innovative xApp design based on the combination of an autoencoder with the \gls{drl} agent to improve the resilience and robustness to real, imperfect network telemetry. We then utilize \coloran to provide insights on the performance of the \gls{drl} agents for adaptive \gls{ran} control at scale. 
We train the autoencoders and agents over a~$3.4$\:GB dataset with more than~73 hours of live \gls{ran} performance traces, and perform one of the first evaluations of \gls{drl} agents autonomously driving a programmable, software-defined \gls{ran} with~49 nodes. 
Lessons learned from this evaluation span from the design to the deployment of \gls{drl} agents for \gls{ran} control. They include new understandings of data analysis and feature selection, modeling of control actions for \gls{drl} agents, and design strategies to train \gls{ml} algorithms that generalize and operate even with unreliable data.

\smallskip
\noindent $\bullet$ We analyze the tradeoffs of training of \gls{drl} agents on live networks using Colosseum and Arena (a publicly-available indoor testbed for spectrum research~\cite{bertizzolo2020arena}) with commercial smartphones.
We profile the \gls{ran} performance during the \gls{drl} exploration phase and after the training, showing how an extra online training step adapts a pre-trained model to deployment-specific parameters, fine-tuning its weights
at the cost of a temporary performance degradation in the online exploration phase.

\smallskip

Key takeaways from our work highlight (i) the effectiveness of adaptive control policies over static configurations, even if the latter are optimized; (ii) the impact of different design choices of \gls{drl} agents on end-to-end network performance, and (iii) the importance of online \gls{drl} training in wireless environments. 
We believe that these insights and the research infrastructure developed in this work can catalyze, promote and further the deployment of \gls{ml}-enabled control loops in next generation networks. 

The rest of this paper is organized as follows. 
Section~\ref{sec:machine-pipelines} describes the development of \gls{ml} solutions in O-RAN-based networks. 
Section~\ref{sec:coloran} introduces \coloran, and Section~\ref{sec:xapps} presents the xApp, \gls{drl} agent design, and the data collection campaign for offline training.  
Large-scale evaluation and lessons learned are discussed in Sections~\ref{sec:results-xapps} and~\ref{sec:online}. 
%
Section~\ref{sec:related} reviews related work. 
Finally, Section~\ref{sec:conclusions} concludes the paper. 

\section{Machine Learning for the Open RAN}
\label{sec:machine-pipelines}

\begin{figure}[ht]
    \centering
    \includegraphics[width=.98\columnwidth]{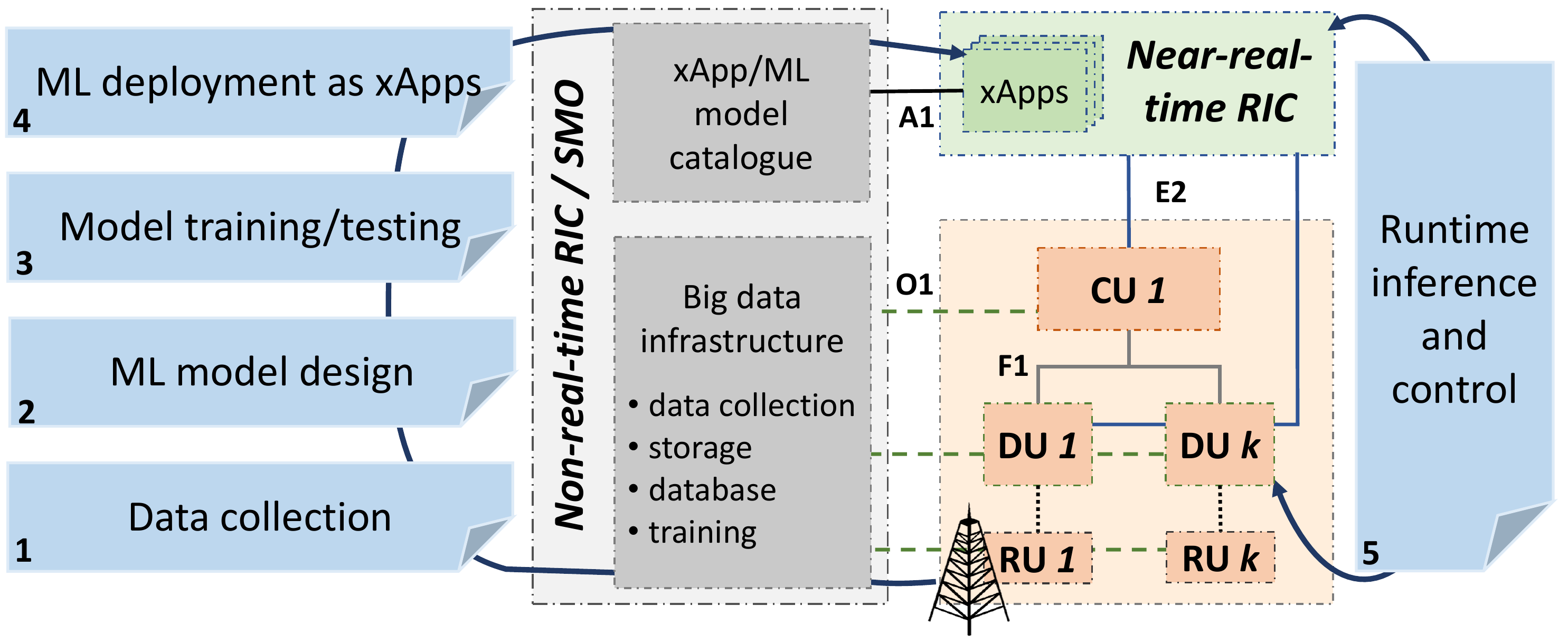}
    \caption{The O-RAN architecture and the workflow for the design, development and deployment of ML applications in next generation wireless networks.}
    \label{fig:oran-primer}
\end{figure}

The deployment of machine learning models in wireless networks is a multi-step process (Fig.~\ref{fig:oran-primer}).
It involves a data collection step, the design of the model, its offline or online training and
deployment for runtime inference and control.
The O-RAN architecture, also shown in Fig.~\ref{fig:oran-primer}, has been developed to aid the overall deployment process, focusing 
on open interfaces for data collection and deployment steps. 
In the following, we  describe the O-RAN architecture, and discuss how it facilitates training and deploying \gls{ml} models in the \gls{ran}.

\subsection{O-RAN Overview}


The O-RAN Alliance, a consortium of academic and industry members, has been pushing forward the concept of an open and programmable cellular ecosystem since its inception in 2018.
%
O-RAN-compliant equipment is based on open and standardized interfaces that enable interoperability of equipment from different vendors and interaction with \gls{ran} controllers, which
manage the \gls{ran} itself.
%
The O-RAN specifications introduce two \glspl{ric} that perform network control procedures over different time scales, i.e.,  \textit{near-real-time} and \textit{non-real-time}, respectively~\cite{oran-arch-spec}.
The non-real-time \gls{ric} performs operations at time scales larger than 1 s and can involve thousands of devices. 
Examples include \gls{smo}, policy management, training and deployment of \gls{ml} models.
%
The near-real-time \gls{ric}, instead, implements tight control loops that span from 10 ms to 1 s, involving hundreds of \glspl{cu}/\glspl{du}.
Procedures for load balancing, handover, \gls{ran} slicing policies~\cite{doro2021coordinated} and scheduler configuration are examples of near-real-time \gls{ric} operations~\cite{oranwpusecases}. 
The near-real-time \gls{ric} can also hosts third-party applications, i.e., \textit{xApps}.
%
xApps implement control logic through heuristics or data-driven control loops, as well as collect and analyze data from the \gls{ran}.


The components of the O-RAN architecture are connected via
open and standardized interfaces. The non-real-time \gls{ric} uses the O1 interface to collect data in bulk from \gls{ran} nodes and to provision services and network functions. 
The near-real-time \gls{ric} connects to \glspl{cu} and \glspl{du} through the E2 interface, which supports different \glspl{sm}, i.e., functionalities like reporting of \glspl{kpm} from \gls{ran} nodes and the control of their parameters~\cite{oran-e2sm-kpm}. 
The two \glspl{ric} connect through the A1 interface for the deployment of policies and xApps on the near-real-time~\gls{ric}.

%


\subsection{ML Pipelines in O-RAN}
\label{sec:ml-pipeline}

The O-RAN specifications include guidelines for the management of \gls{ml} models in cellular networks.
Use cases and applications include \gls{qos} optimization and prediction, traffic steering, handover, and radio fingerprinting~\cite{oran-ml}.
The specifications describe the \gls{ml} workflow for O-RAN through five steps (Fig.~\ref{fig:oran-primer}):~(1)~data collection; (2)~model design; (3)~model training and testing; (4)~model deployment as xApp, and~(5)~runtime inference and control.

%
First, data is collected for different configurations and setups of the \gls{ran} (e.g., large/small scale, different traffic, step~1). Data is generated by the \gls{ran} nodes, i.e., \glspl{cu}, \glspl{du} and \glspl{ru}, and streamed to the non-real-time \gls{ric} through the O1 interface, where it is organized in large datasets.
After enough data has been collected, an \gls{ml} model is designed (step~2). This entails the following: (i) identifying the \gls{ran} parameters to input to the model (e.g., throughput, latency, etc.); (ii) identifying the \gls{ran} parameters to control as output (e.g., \gls{ran} slicing and scheduling policies), and (iii) the actual \gls{ml} algorithm implementation.
Once the model has been designed and implemented, it is trained and tested on the collected data (step~3). This involves selecting the model hyperparameters (e.g., the depth and number of layers of the neural network) and training the model on a portion of the collected data until a (satisfactory) level of convergence of the model has been reached. 
After the model has been trained, it is tested on an unseen portion of the collected data to verify that it is able to generalize and react to potentially unforeseen situations.
%
%
Then, the model is packaged into an xApp ready to run on the \mbox{near-real-time \gls{ric}} (step~4).
%
After the xApp has been created, it is deployed on the O-RAN infrastructure. In this phase, the model is first stored in the xApp catalogue of the non-real-time \gls{ric}, and then instantiated on demand on the near-real-time \gls{ric}, where it
%
is interfaced with the \gls{ran}  through the E2 interface to perform runtime inference and control based on the current network conditions (step~5).

\begin{figure*}[t]
    \centering
    \includegraphics[width=\textwidth]{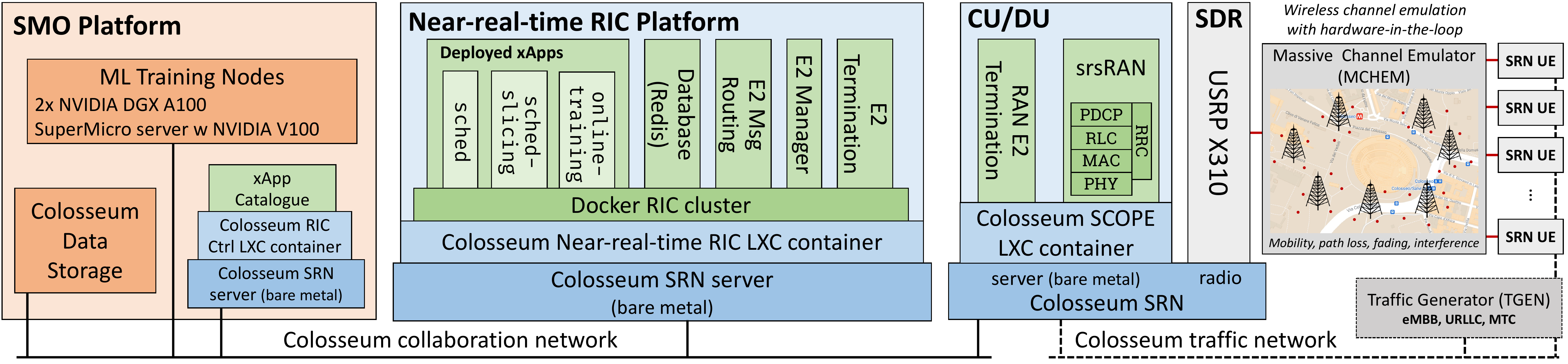}
    \caption{Integration of the O-RAN infrastructure in Colosseum.}
    \label{fig:oran-colosseum}
\end{figure*}

\section{\coloran: Enabling Large-scale \gls{ml} Research with O-RAN and Colosseum}
\label{sec:coloran}

The ML pipeline described in Section~\ref{sec:ml-pipeline} involves a number of critical steps whose execution requires joint access to \emph{comprehensive datasets} and \emph{testing facilities at scale}, still largely unavailable to the research community.
In fact, even major telecom operators or infrastructure owners might not be able to dedicate (parts of) their extensive commercial networks to training and testing of \gls{ml} algorithms. This stems from the lack of adequate solutions to separate testing from commercial service and to prevent performance degradation.
As a consequence, researchers and innovators are constrained to work with small ad hoc datasets collected in contained lab setups, resulting in solutions that hardly generalize to real-world deployments~\cite{wang2020}.

To address this limitation, this section introduces \coloran, a large-scale research infrastructure built upon the Colosseum network emulator to train, deploy, and test state-of-the-art wireless \gls{ml} solutions.
We first review the main features of Colosseum and describe its use as a wireless data factory for \coloran (Section~\ref{sec:df}).
Then, we introduce the implementation of the \coloran virtualized O-RAN infrastructure on Colosseum (Section~\ref{sec:oran}) and of the xApps we designed (Section~\ref{sec:xapps}).
We finally describe the scenario for data collection that we use to illustrate the usage of \coloran (Section~\ref{sec:data}).

\subsection{Colosseum as a Wireless Data Factory}
\label{sec:df}

Colosseum is the world's largest wireless network emulator~\cite{bonati2021colosseum}.
It was developed by DARPA for the Spectrum Collaboration Challenge and then transitioned to the U.S.\ National Science Foundation PAWR program to be available for
the research community. Colosseum includes~256 USRP X310 \glspl{sdr}.
Half of the \glspl{sdr} can be controlled by the users, while the other half is part of the \gls{mchem}, which uses 64 Virtex-7 \acrshort{fpga}s to emulate wireless channels. 
\gls{mchem} processes the signals transmitted by radio nodes---called \glspl{srn} in Colosseum---through a set of complex-valued finite impulse response filter banks. 
These
model propagation characteristics and multi-path scattering of user-defined wireless environments, as shown in the right part of Fig.~\ref{fig:oran-colosseum}. 
%
Thus, \gls{mchem} provides high-fidelity emulation of wireless signals with the same characteristics of those traveling through a real environment.
Colosseum also features a user-controlled source \gls{tgen}, based on \acrshort{mgen}~\cite{mgen}, and compute capabilities that make it a full-fledged specialized data center with over~$170$ high-performance servers.

The combination of programmable software-defined hardware with \gls{rf} and traffic scenarios uniquely positions Colosseum as a wireless data factory, namely, as a tool that can be used to effectively collect full-stack datasets in heterogeneous and diverse scenarios.
With respect to other large testbeds such as the PAWR platforms, Colosseum offers scale and a more controlled and customizable environment that researchers can use to collect data and to test \gls{ml} algorithms on different \gls{rf} scenarios and frequencies, without changing the protocol stack or experimental procedures. 
Compared to a production network, Colosseum is flexible, with programmable radios that can run different software-defined stacks, and the possibility to test closed-loop control without affecting commercial deployments.

\subsection{O-RAN-based Colosseum ML Infrastructure}
\label{sec:oran}

Besides enabling large-scale data collection, Colosseum also provides a hybrid \gls{rf} and compute environment for the deployment of \coloran, a complete end-to-end \gls{ml} infrastructure.
%
%
\coloran provides researchers with a ready-to-use environment to develop and test \gls{ml} solutions, following the steps of Fig.~\ref{fig:oran-primer} (Section~\ref{sec:ml-pipeline}).
These include the deployment on a \gls{3gpp}-compliant \gls{ran}, testing in heterogeneous emulated environments, and an O-RAN-compliant infrastructure.
With respect to other open source implementations of the O-RAN infrastructure, \coloran features a more lightweight footprint (e.g., it does not require a full Kubernetes deployment, contrary to the \gls{osc} \gls{ric}), and it can be ported to other testbeds, e.g., Arena~\cite{bertizzolo2020arena},
with minimal changes, thanks to its virtualized and container-based implementation.
As a further contribution, this platform will be made openly available to the research community upon acceptance of this paper.

The software, compute and networking components of our end-to-end infrastructure are shown in Fig.~\ref{fig:oran-colosseum}.
%
The \gls{smo} (left) features three compute nodes to train large \gls{ml} models,
$64$~Terabyte of storage for models and datasets, and the xApp catalogue.
%
The near-real-time \gls{ric} (Fig.~\ref{fig:oran-colosseum}, center) provides E2 connectivity to the \gls{ran} and support for multiple xApps interacting with the base stations. It is implemented as a standalone \gls{lxc} that can be deployed on a Colosseum \gls{srn}.
%
It includes multiple Docker containers for the \textit{E2 termination} and \textit{manager}, the \textit{E2 message routing} to handle messages internal to the \gls{ric}, a \textit{Redis database}, which keeps a record of the nodes connected to the \gls{ric}, and the \textit{xApps} (Section~\ref{sec:xapps}). 
The implementation of the near-real-time \gls{ric} is based on the Bronze release of the \gls{osc}~\cite{bronze_release}. The \gls{osc} near-real-time \gls{ric} was adapted into a minimal version, which does not require a Kubernetes cluster, and can fit in a lightweight LXC container. We also extended the \gls{osc} codebase to support concurrent connections from multiple base stations and xApps, and to provide improved support for encoding, decoding and routing of control messages.

The near-real-time \gls{ric} connects to the \gls{ran} base stations through the E2 interface (Fig.~\ref{fig:oran-colosseum}, right).
The base stations leverage a joint implementation of the \gls{3gpp} \glspl{du} and \glspl{cu}.
These nodes run the publicly available SCOPE framework~\cite{bonati2021scope}, which extends srsRAN~\cite{gomez2016srslte} with open interfaces for runtime reconfiguration of base station parameters and automatic collection of relevant
\glspl{kpm}.
Moreover, we leverage and extend the E2 termination of the \gls{osc} \gls{du}~\cite{oran_du} to reconfigure the base stations directly from the near-real-time \gls{ric} and for periodic data reporting.
%
The E2 termination allows the setup procedure and registration of the base stations with the near-real-time \gls{ric}.
Our implementation also features two custom \glspl{sm} (as discussed next) for trigger-based or periodic reporting, and control events in the base stations. This effectively enables data-driven real-time control loops between the base stations and the xApps. The \gls{ran} supports network slicing with~3 slices for different \gls{qos}: (i)~\gls{embb}, representing users requesting video traffic;
(ii)~\gls{mtc} for sensing applications, and (iii)~\gls{urllc} for latency-constrained applications. Slicing is implemented in the SCOPE framework by applying \gls{prb} masks during the scheduling process, and it is possible to control the number of \glspl{prb} for each slice~\cite{bonati2021scope}.
For each slice, the base stations can adopt~3 different scheduling policies independently of that of the other slices, namely, the \gls{rr}, the \gls{wf}, and the \gls{pf} scheduling policies. These policies were selected as they represent popular scheduling strategies in wireless deployments~\cite{capozzi2013downlink}.
%
Finally, the base stations connect to the \gls{rf} frontends (USRPs X310)
that perform signal transmission and reception.


\section{xApp Design for \gls{drl}-based Control}
\label{sec:xapps}


\begin{table*}[t]
    \centering
    \footnotesize
    \setlength{\tabcolsep}{2pt}
    \begin{tabularx}{\linewidth}{>{\hsize=.29\hsize}X>{\raggedright\arraybackslash\hsize=.74\hsize}X>{\raggedright\arraybackslash\hsize=.49\hsize}X>{\raggedright\arraybackslash\hsize=.53\hsize}X>{\raggedright\arraybackslash\hsize=.74\hsize}X>{\raggedright\arraybackslash\hsize=\hsize}X}
        \toprule
         xApp & Functionality & Input (Observation) & Output (Action) & \gls{ml} Models & Utility (Reward) \\\midrule
         \texttt{sched-} \texttt{slicing} & Single-\gls{drl}-agent for joint slicing and scheduling control & Rate, buffer size, PHY TBs (DL) & \gls{prb} and scheduling policy for each slice & DRL-base, DRL-reduced- actions, DRL-no-autoencoder & Maximize rate for \gls{embb}, PHY TBs for \gls{mtc}, minimize buffer size for \gls{urllc}\\
         \rule{0pt}{2.5ex}\texttt{sched} & Multi-\gls{drl}-agent per-slice scheduling policy selection & Rate, buffer size, \gls{prb} ratio (DL) & Scheduling policy for each slice & \gls{drl}-sched & Maximize rate for \gls{embb} and \gls{mtc}, \gls{prb} ratio for \gls{urllc} \\
         \rule{0pt}{2.5ex}\texttt{online-} \texttt{training} & Train \gls{drl} agents with online exploration & Rate, buffer size, PHY TBs (DL) & Training action (\gls{prb} and scheduling) & Trained online by the xApp itself & Based on specific training goals \\
         \bottomrule
    \end{tabularx}
    \caption{Catalogue of developed xApps.}
    \label{tab:xapps}
\end{table*}

The xApps deployed on the near-real-time \gls{ric} are the heart of the O-RAN-based \gls{ran} control loops. We developed three xApps to evaluate the impact of different \gls{ml} strategies for closed-loop \gls{ran} control (Table~\ref{tab:xapps}). Each xApp can receive data and control \gls{ran} nodes with two custom \glspl{sm}, which resemble the O-RAN \gls{kpm} and \gls{ran} control \glspl{sm}~\cite{oran-e2sm-kpm}. The control actions available to the xApps are the selection of the slicing policy (the number of \gls{prb} allocated to each slice)
and of the scheduling policy (which scheduler
is used for each slice).

\begin{figure}[h]
    \centering
    \includegraphics[width=\columnwidth]{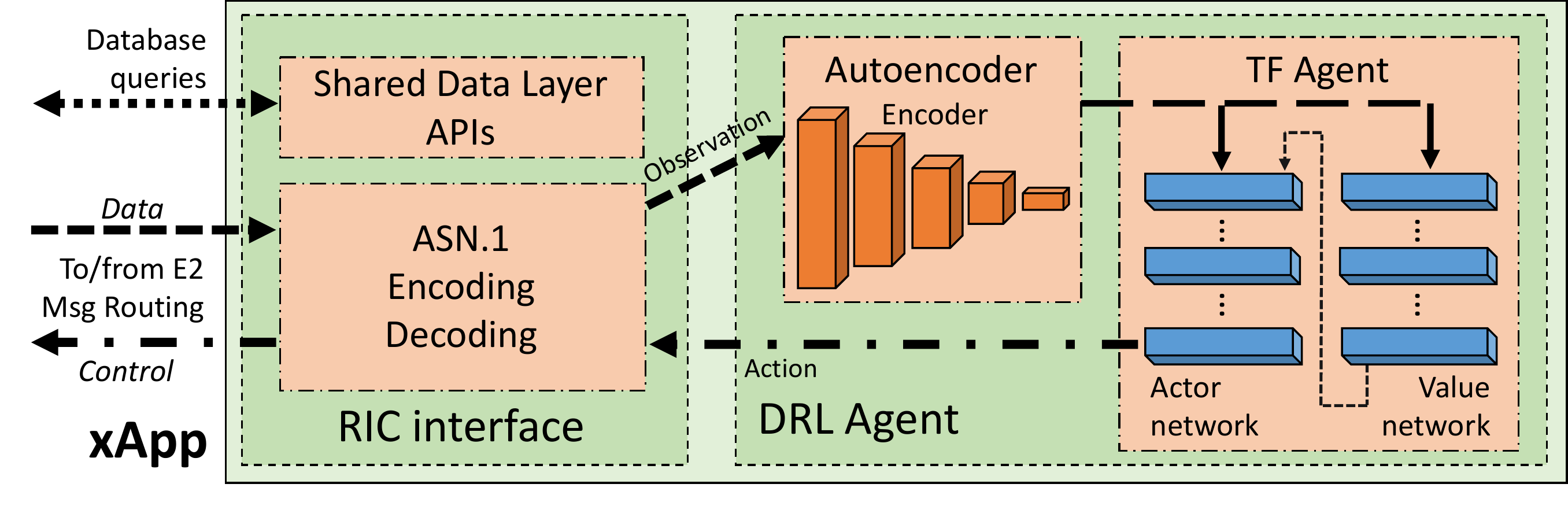}
    \caption{Structure of a \coloran xApp.}
    \label{fig:xapp}
\end{figure}

The xApps have been developed by extending the \gls{osc} basic xApp framework~\cite{oran_xapp}, and include two components (Fig.~\ref{fig:xapp}).

\noindent The first is the interface to the \gls{ric}, which implements the \gls{sm} and performs
ASN.1 encoding/decoding of \gls{ran} data and control. The second is the \gls{ml} infrastructure itself, which includes one or more autoencoders and \gls{drl} agents. For these, we used TensorFlow 2.4~\cite{tensorflow2015-whitepaper} and the TF-Agents library~\cite{TFAgents}.

\subsection{\gls{drl} agent design}

The \gls{drl} agents considered in this paper have been trained using the \gls{ppo} algorithm~\cite{schulman2017proximal}.
\gls{ppo} is a well-established on-policy \gls{drl} architecture that uses an actor-critic configuration where the \textit{actor} network takes actions according to current network state, and the value network (or \textit{critic}) scores the actions taken by the actor network by observing the reward obtained when taking an action in a specific state of the environment. By leveraging this architecture, the \gls{ppo} algorithm decouples the action taking process from the evaluation of achieved rewards. This is extremely important to ensure that the actor network can learn an unbiased policy (i.e., a mapping between state and actions) where the actor network selects an action because it is effective in the long run and not only because it occasionally results in high instantaneous rewards that are instead inefficient in the majority of cases. 

It is also worth mentioning that the actor-critic setup is also important because \gls{ppo} is an on-policy architecture, which means that the training procedure uses a memory buffer that contains data that is collected by using actions that are taken with the most current version of the actor network. If compared to off-policy algorithms (such as \glspl{dqn}), which use a memory buffer that store experience collected at any time by the \gls{drl} agent, \gls{ppo} only uses data that is fresh and not contain experiences from the past, meaning that the memory buffer is emptied every time the actor network is updated during the training phase. This approach is usually slower then other solutions, but together with the actor-critic setup it has been demonstrated as one of the most efficient and reliable \gls{drl} architectures in the literature~\cite{schulman2017proximal}. 

\textbf{Observations, Actions and Rewards. } One of the main causes of slow training of \gls{drl} agents is the use of observations with high dimensionality that result in actor and critic networks with many parameters and large state space. Indeed, the RAN produces an extremely large amount of data which not always provide meaningful insights on the actual state of the system due to redundant information and outliers. To reduce the size of the observation fed to the \gls{drl} agent, mitigate outliers and provide a high-quality yet high-level representation of the state of the system, we resort to autoencoders, as also shown in Fig.~\ref{fig:xapp}. Specifically, before being fed to the \gls{drl} agents, the data produced by the RAN is processed by the encoding portion of an autoencoder for dimensionality reduction (whose impact on \gls{drl}-based control is investigated in Section~\ref{sec:models}).

Although autoencoders might have several implementations according to the specific applications, autoencoders for dimensionality reduction have an hourglass architecture with an encoder and a decoder components. The former produces a lower dimension representation of the input data (i.e., latent representation) which - if trained properly - can be accurately reconstructed by the decoder portion of the autoencoder with negligible error. The decoder is the specular image of the encoder and the goal of this architecture is to create a reduced version of the input data that contains only relevant information, yet it is accurate enough to be able to reconstruct the original data without any loss.  
To further reduce the complexity of the \gls{drl} agents, we perform feature selection on the metrics that are observed by the agents (see Section \ref{sec:results-xapps} for more details).

As mentioned before, each xApp embeds different \gls{drl} agents according to the specific goal of the xApp. For this reason, we have designed a set of \gls{drl} agents that observe different metrics of the RAN, take diverse actions and aim at maximizing different rewards. The configurations considered in this paper are shown in Table~\ref{tab:xapps}.
The \gls{drl} agent of \texttt{sched-slicing} jointly selects the slicing and scheduling policy for a single base station and all slices. For this xApp we trained three \gls{drl} models: baseline (\gls{drl}-base), an agent that explores a reduced set of actions (\gls{drl}-reduced-actions) and an agent where input data 
is fed directly to the agent (\gls{drl}-no-autoencoder).
The \texttt{sched} xApp includes three \gls{drl} agents that select in parallel the scheduling policy for each slice (\gls{embb}, \gls{mtc}, and \gls{urllc}). Each agent has been trained using slice-specific data.

\subsection{Training the \gls{drl} Agents}

\gls{drl} agents are trained on the dataset described in Section~\ref{sec:data}, where at each training episode we select \gls{ran} data from different base stations to remove dependence on a specific wireless environment (Section~\ref{sec:online}) and facilitate generalization.

Following O-RAN specifications, training is performed offline on the dataset. In our case, this is achieved by randomly selecting instances in which the network reaches the state~$s_1$ that results from the combination of the previous state~$s_0$ and the action to explore~$a_0$. 

In our experiments, the actor and critic networks of all \gls{drl} agents have been implemented as two fully-connected neural networks with~5 layers with~30 neurons each and an hyperbolic tangent activation function. 
The encoder consists of~4 fully-connected layers with~256, 128, 32 and~3 neurons and a rectified linear activation function. 
For all models, the learning rate is set to~$0.001$. 

Finally, as illustrated in Table~\ref{tab:xapps}, we also consider the case of online training where the \texttt{online-training} xApp supports training a \gls{drl} agent using live data from the \gls{ran} and performing exploration steps on the online \gls{ran} infrastructure itself. While this is not recommended by
O-RAN~\cite{oran-ml}, it
specializes
the trained model to the specific
deployment. We will discuss the tradeoffs involved in this operation in Section~\ref{sec:online}. \texttt{online-training} leverages TensorFlow \texttt{CheckPoint} objects to save and restore a (partially) trained model for multiple consecutive rounds of training. In this way, the training services in the 
xApp can restore an agent trained on an offline dataset using it as starting point for the online, live training on the \gls{ran}.

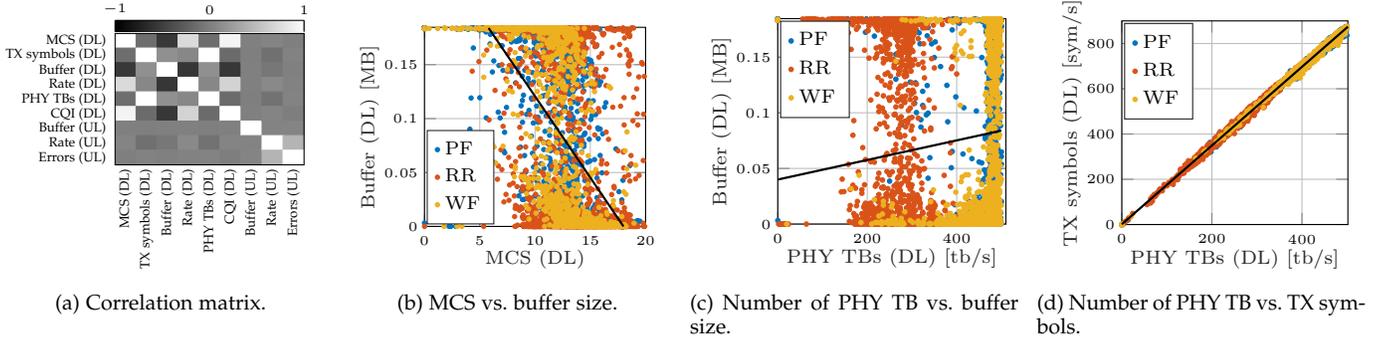
\begin{figure*}[t]
\begin{subfigure}[t]{0.24\textwidth}
	\centering
	\setlength\fwidth{.575\columnwidth}
	\setlength\fheight{.4\columnwidth}
	\begin{tikzpicture}
\pgfplotsset{every tick label/.append style={font=\tiny}}

\begin{axis}[%
width=\fwidth,
height=\fheight,
at={(0\fwidth,0\fheight)},
scale only axis,
point meta min=-1,
point meta max=1,
axis on top,
xtick=data,
ytick=data,
xmin=-0.5,
xmax=8.5,
xticklabels={MCS (DL), TX symbols (DL), Buffer (DL), Rate (DL), PHY TBs (DL), CQI (DL), Buffer (UL), Rate (UL), Errors (UL)},
y dir=reverse,
ymin=-0.5,
ymax=8.5,
yticklabels={MCS (DL), TX symbols (DL), Buffer (DL), Rate (DL), PHY TBs (DL), CQI (DL), Buffer (UL), Rate (UL), Errors (UL)},
xlabel style={font=\footnotesize\color{white!15!black}},
ylabel style={font=\footnotesize\color{white!15!black}},
axis background/.style={fill=white},
colormap/blackwhite,
colorbar,
enlargelimits=false,
scale only axis,
tick align=inside,
xtick style={draw=none},
ytick style={draw=none},
colorbar horizontal,
colorbar style={
at={(0,1.01)},               
anchor=below south west,    
width=\pgfkeysvalueof{/pgfplots/parent axis width},
xtick={-1, 0, 1},
xmin=-1,
xmax=1,
axis x line*=top,
xticklabel shift=-1pt,
point meta min=-1,
point meta max=1,
},
colorbar/width=1.5mm,
xticklabel style={rotate=90},
xticklabel shift=-2pt,
]
\addplot [matrix plot,point meta=explicit]
 coordinates {
(0,0) [1] (0,1) [-0.14158] (0,2) [-0.56038] (0,3) [0.70155] (0,4) [-0.14103] (0,5) [0.89749] (0,6) [-0.0016836] (0,7) [0.025515] (0,8) [-0.0127] 

(1,0) [-0.14158] (1,1) [1] (1,2) [0.12403] (1,3) [0.049395] (1,4) [0.99895] (1,5) [-0.16612] (1,6) [-0.0083726] (1,7) [-0.12501] (1,8) [0.011295] 

(2,0) [-0.56038] (2,1) [0.12403] (2,2) [1] (2,3) [-0.58301] (2,4) [0.12442] (2,5) [-0.56955] (2,6) [-0.0033958] (2,7) [-0.075853] (2,8) [-0.0027763] 

(3,0) [0.70155] (3,1) [0.049395] (3,2) [-0.58301] (3,3) [1] (3,4) [0.050409] (3,5) [0.64218] (3,6) [0.00064405] (3,7) [0.066629] (3,8) [0.0087001] 

(4,0) [-0.14103] (4,1) [0.99895] (4,2) [0.12442] (4,3) [0.050409] (4,4) [1] (4,5) [-0.16473] (4,6) [-0.0079734] (4,7) [-0.11358] (4,8) [0.023103] 

(5,0) [0.89749] (5,1) [-0.16612] (5,2) [-0.56955] (5,3) [0.64218] (5,4) [-0.16473] (5,5) [1] (5,6) [-0.0032019] (5,7) [0.028991] (5,8) [-0.015027] 

(6,0) [-0.0016836] (6,1) [-0.0083726] (6,2) [-0.0033958] (6,3) [0.00064405] (6,4) [-0.0079734] (6,5) [-0.0032019] (6,6) [1] (6,7) [0.021668] (6,8) [0.027198] 

(7,0) [0.025515] (7,1) [-0.12501] (7,2) [-0.075853] (7,3) [0.066629] (7,4) [-0.11358] (7,5) [0.028991] (7,6) [0.021668] (7,7) [1] (7,8) [0.40338] 

(8,0) [-0.0127] (8,1) [0.011295] (8,2) [-0.0027763] (8,3) [0.0087001] (8,4) [0.023103] (8,5) [-0.015027] (8,6) [0.027198] (8,7) [0.40338] (8,8) [1] 

};
\end{axis}
\end{tikzpicture}
	\caption{Correlation matrix.}
	\label{fig:corr-matrix-embb}
\end{subfigure}\hfill
\begin{subfigure}[t]{0.24\textwidth}
	\centering
	\includegraphics[width=.95\textwidth]{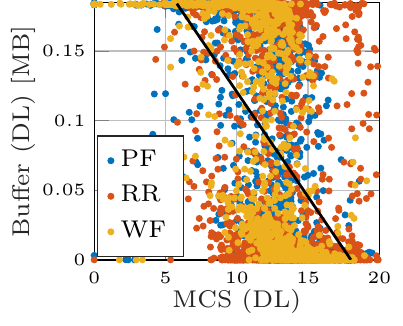}
	\caption{MCS vs.\ buffer size.}
	\label{fig:mcs-buf-embb}
\end{subfigure}\hfill
\begin{subfigure}[t]{0.24\textwidth}
	\centering
	\includegraphics[width=.925\textwidth]{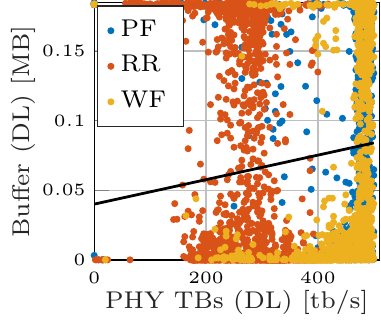}
	\caption{Number of PHY \acrshort{tb} vs.\ buffer size.}
	\label{fig:pkt-buf-embb}
\end{subfigure}\hfill
\begin{subfigure}[t]{0.24\textwidth}
	\centering
	\includegraphics[width=.9\textwidth]{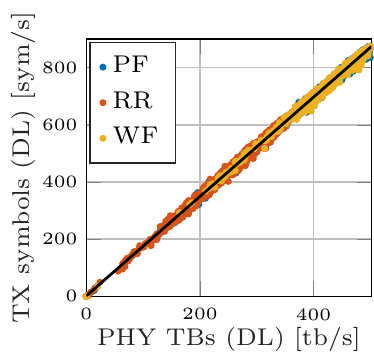}
	\caption{Number of PHY \acrshort{tb} vs.\ TX symbols.}
	\label{fig:pkt-n-samples-embb}
\end{subfigure}
\caption{Correlation analysis for the eMBB slice with 36 PRBs and the slice-based traffic profile. The solid line is the linear regression fit of the data.}
\label{fig:corr-embb0}
\end{figure*}

\subsection{Large-scale Data Collection for \coloran}
\label{sec:data}


To train the \gls{drl} agents for the \coloran xApps we performed large-scale data collection experiments on Colosseum. The parameters for the scenario are summarized in Table~\ref{tab:scenario}.

\begin{table}[t]
    \centering
    \footnotesize
    \setlength{\tabcolsep}{2pt}
    \begin{tabularx}{\linewidth}{>{\hsize=.31\hsize}X>{\hsize=1\hsize}X}
        \toprule
         Parameter & Value \\ \midrule
         Number~of nodes & $N_{\rm BS} \!=\! 7$, $N_{\rm UE} \!=\! 42$ \\
         \rule{0pt}{2.5ex}RF parameters & DL carrier $f_d \!=\! 0.98$ GHz, UL carrier $f_u \!=\! 1.02$ GHz, bandwidth $B \!=\! 10$ MHz (50 PRBs) \\
         \rule{0pt}{2.5ex}Schedulers & \acrshort{rr}, \acrshort{wf}, \acrshort{pf} \\
         \rule{0pt}{2.5ex}Slices & \acrshort{embb}, \acrshort{mtc}, \acrshort{urllc} (2 UEs/BS/slice) \\
         \rule{0pt}{2.5ex}Traffic profiles & Slice-based: $4$ Mbit/s/UE for eMBB, $44.6$ kbit/s/UE for MTC, $89.3$ kbit/s/UE URLLC \\
         & Uniform: $1.5$ Mbit/s/UE for eMBB, MTC, URLLC \\
         \bottomrule
    \end{tabularx}
    \caption{Configuration parameters for the considered scenario.}
    \label{tab:scenario}
\end{table}

The large-scale \gls{rf} scenario mimics a real-world cellular deployment in downtown Rome, Italy, with the positions of the base stations derived from the OpenCelliD database~\cite{opencellid}.
We instantiated a softwarized cellular network with 7 base stations through the SCOPE framework. Each base station operates on a $10$~MHz channel
($50$ \glspl{prb}) which can be dynamically assigned to the 3 slices (i.e., \gls{embb}, \gls{mtc}, \gls{urllc}).
Additionally, we considered two different \gls{tgen} traffic scenarios: slice-based traffic and uniform traffic.
In slice-based traffic,
users are distributed among
different traffic profiles ($4$~Mbit/s constant bitrate traffic to \gls{embb} users,
and $44.6$~kbit/s and $89.3$~kbit/s Poisson traffic to \gls{mtc} and \gls{urllc}, respectively).
%
The uniform traffic is configured with~$1.5$~Mbit/s for all users. The training of the \gls{drl} agents on the offline dataset has been performed with slice-based traffic.
Finally, the base stations serve a total of 42 users equally divided among the 3 slices.
%

In our data collection campaign, we gathered $3.4$\:GB of data, for a total of more than 73~hours of experiments.
In each experiment, the base stations periodically report \gls{ran} \glspl{kpm} to the
non-real-time \gls{ric}. These include metrics such as throughput, buffer queues, number of PHY \glspl{tb} and \glspl{prb}. The complete dataset features more than 30 metrics that can be used for \gls{ran} analysis and \gls{ml} training.

\section{\gls{drl}-based xApp Evaluation}
\label{sec:results-xapps}

Learning strategies for \gls{ran} control are coded as xApps on \coloran.
This section presents their comparative performance evaluation.
Feature selection based on \gls{ran} \glspl{kpm} is described in Section~\ref{sec:features}.
The experimental comparison of the different \gls{drl} models is reported in Section~\ref{sec:models}.

\subsection{\gls{ran} \gls{kpm} and Feature Selection}
\label{sec:features}

O-RAN is the first architecture
to introduce a standardized way to extract telemetry and data from the \gls{ran} to drive closed-loop control. 
However, 
O-RAN does not indicate which \glspl{kpm} should be considered for the design of ML algorithms. 
%
The O-RAN E2SM \gls{kpm} specifications~\cite{oran-e2sm-kpm} allow the generation of more than 400 possible \glspl{kpm}, listed in~\cite{3gpp.28.552,3gpp.32.425}. More vendor-specific \glspl{kpm} may also be reported on E2.
These \glspl{kpm} range from physical layer metrics to base station monitoring statistics. Therefore, the bulk set of data may not be useful to represent the network state for a specific problem. Additionally, reporting or collecting all the metrics via the E2 or O1 interfaces introduces a high overhead, and a highly dimensional input may lead to sub-optimal performance for \gls{ml}-driven xApps~\cite{sakurada2014anomaly}.

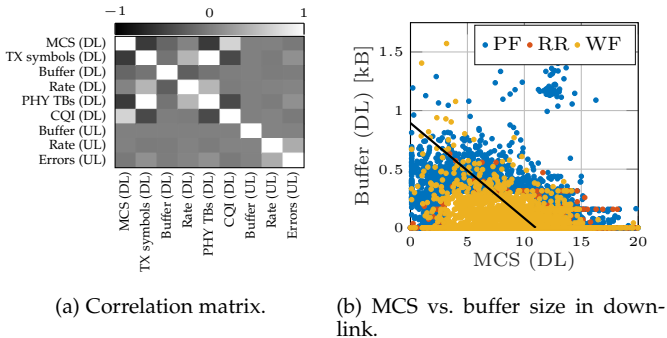
\begin{figure}[b]
\begin{subfigure}[t]{0.24\textwidth}
	\centering
	\setlength\fwidth{.575\columnwidth}
	\setlength\fheight{.4\columnwidth}
	\begin{tikzpicture}
\pgfplotsset{every tick label/.append style={font=\tiny}}

\begin{axis}[%
width=\fwidth,
height=\fheight,
at={(0\fwidth,0\fheight)},
scale only axis,
point meta min=-1,
point meta max=1,
axis on top,
xtick=data,
ytick=data,
xmin=-0.5,
xmax=8.5,
xticklabels={MCS (DL), TX symbols (DL), Buffer (DL), Rate (DL), PHY TBs (DL), CQI (DL), Buffer (UL), Rate (UL), Errors (UL)},
y dir=reverse,
ymin=-0.5,
ymax=8.5,
yticklabels={MCS (DL), TX symbols (DL), Buffer (DL), Rate (DL), PHY TBs (DL), CQI (DL), Buffer (UL), Rate (UL), Errors (UL)},
xlabel style={font=\footnotesize\color{white!15!black}},
ylabel style={font=\footnotesize\color{white!15!black}},
axis background/.style={fill=white},
colormap/blackwhite,
colorbar,
enlargelimits=false,
scale only axis,
tick align=inside,
xtick style={draw=none},
ytick style={draw=none},
colorbar horizontal,
colorbar style={
at={(0,1.01)},               
anchor=below south west,    
width=\pgfkeysvalueof{/pgfplots/parent axis width},
xtick={-1, 0, 1},
xmin=-1,
xmax=1,
axis x line*=top,
xticklabel shift=-1pt,
point meta min=-1,
point meta max=1,
},
colorbar/width=1.5mm,
xticklabel style={rotate=90},
xticklabel shift=-2pt
]
\addplot [matrix plot,point meta=explicit]
 coordinates {
(0,0) [1] (0,1) [-0.54841] (0,2) [-0.20418] (0,3) [0.022546] (0,4) [-0.54612] (0,5) [0.64062] (0,6) [-0.00045741] (0,7) [0.0068991] (0,8) [-0.061182] 

(1,0) [-0.54841] (1,1) [1] (1,2) [-0.097322] (1,3) [0.40597] (1,4) [0.99472] (1,5) [-0.43452] (1,6) [-0.0018217] (1,7) [-0.039358] (1,8) [0.11823] 

(2,0) [-0.20418] (2,1) [-0.097322] (2,2) [1] (2,3) [-0.2591] (2,4) [-0.0987] (2,5) [-0.019518] (2,6) [-0.00077487] (2,7) [-0.032014] (2,8) [-0.027517] 

(3,0) [0.022546] (3,1) [0.40597] (3,2) [-0.2591] (3,3) [1] (3,4) [0.41691] (3,5) [-0.030965] (3,6) [-0.000208] (3,7) [0.013455] (3,8) [0.03013] 

(4,0) [-0.54612] (4,1) [0.99472] (4,2) [-0.0987] (4,3) [0.41691] (4,4) [1] (4,5) [-0.42552] (4,6) [-0.0016796] (4,7) [-0.032764] (4,8) [0.12461] 

(5,0) [0.64062] (5,1) [-0.43452] (5,2) [-0.019518] (5,3) [-0.030965] (5,4) [-0.42552] (5,5) [1] (5,6) [-0.00085919] (5,7) [0.022928] (5,8) [-0.052056] 

(6,0) [-0.00045741] (6,1) [-0.0018217] (6,2) [-0.00077487] (6,3) [-0.000208] (6,4) [-0.0016796] (6,5) [-0.00085919] (6,6) [1] (6,7) [0.023903] (6,8) [0.0090925] 

(7,0) [0.0068991] (7,1) [-0.039358] (7,2) [-0.032014] (7,3) [0.013455] (7,4) [-0.032764] (7,5) [0.022928] (7,6) [0.023903] (7,7) [1] (7,8) [0.34298] 

(8,0) [-0.061182] (8,1) [0.11823] (8,2) [-0.027517] (8,3) [0.03013] (8,4) [0.12461] (8,5) [-0.052056] (8,6) [0.0090925] (8,7) [0.34298] (8,8) [1] 

};
\end{axis}
\end{tikzpicture}
	\caption{Correlation matrix.}
	\label{fig:corr-matrix-slice2}
\end{subfigure}\hfill
\begin{subfigure}[t]{0.24\textwidth}
	\centering
	\includegraphics[width=.95\textwidth]{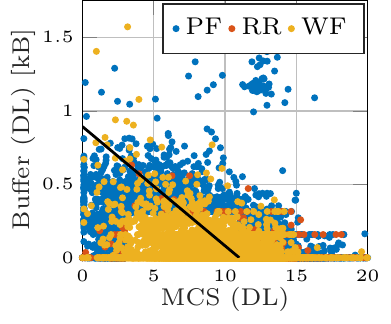}
	\caption{MCS vs.\ buffer size in downlink.}
	\label{fig:mcs-buf-slice2}
\end{subfigure}\hfill
\caption{Correlation analysis for the URLLC slice with 11 PRBs and the slice-based traffic profile. The solid line is the linear regression fit of the data.}
\label{fig:corr-metrics-slice2}
\end{figure}

Therefore, 
a key step in the design process of \gls{ml}-driven xApps is the selection of the features that should be reported for \gls{ran} closed-loop control. In this context, the availability of large-scale, heterogeneous datasets and wireless data factories is key to enable feature selection based on a combined expert- and data-driven approach. To better illustrate this, in Fig.~\ref{fig:corr-embb0} and~\ref{fig:corr-metrics-slice2} we report a correlation analysis for several metrics collected in the dataset described in Section~\ref{sec:data}. The correlation analysis helps us identify the \glspl{kpm} that provide a meaningful description of the network state with minimal redundancy.

\textbf{Correlation analysis}.
Figure~\ref{fig:corr-matrix-embb} shows the correlation matrix of~9 among the~30 UE-specific metrics in the dataset for the \gls{embb} slice.
While downlink and uplink metrics exhibit a low correlation, most downlink \glspl{kpm} positively or negatively correlate with each other (the same holds for uplink \glspl{kpm}). For example, the downlink \gls{mcs} and buffer occupancy have a negative correlation ($-0.56$). This can also be seen in the scatter plot of Fig.~\ref{fig:mcs-buf-embb}: as the \gls{mcs} increases, it is less likely to have a high buffer occupancy, and vice versa. Similarly, the number of \glspl{tb} and symbols in downlink have a strong positive correlation ($0.998$), as also shown in Fig.~\ref{fig:pkt-n-samples-embb}. Two downlink metrics that do not correlate well, instead, are the number of \glspl{tb} and the buffer occupancy. Indeed, the amount of data transmitted in each \gls{tb} varies with the \gls{mcs} and therefore cannot be used as indicator of how much the buffer will empty after each transmission. Additionally, as shown in Fig.~\ref{fig:pkt-buf-embb}, the three scheduling policies have a different quantitative behavior, but they all show a low correlation.  

\textbf{eMBB vs.\ URLLC.} The correlation among metrics also depends on the \gls{ran} configuration and slice traffic profile.
This can be seen by comparing Fig.~\ref{fig:corr-embb0}, which analyzes the \gls{embb} slice with 36 \glspl{prb}, and Fig.~\ref{fig:corr-metrics-slice2}, which uses telemetry for the \gls{urllc} slice with 11 \glspl{prb}. With the slice-based traffic,
the \gls{urllc} users receive data at a rate that is an order of magnitude smaller than that of the \gls{embb} users. As a consequence, the load on the \gls{urllc} slice (represented by the buffer occupancy of Fig.~\ref{fig:mcs-buf-slice2}) is lower, and the buffer is quickly drained even with lower \glspl{mcs}. Consequently, the correlation among the buffer occupancy and the \gls{mcs} ($-0.2$) is lower with respect to the \gls{embb} slice. This further makes the case for collecting datasets that are truly representative of a wireless \gls{ran} deployment, including heterogeneous traffic and diverse applications. 

\textbf{Summary.} Figure~\ref{fig:corr-embb0} and~\ref{fig:corr-metrics-slice2} provide insights on which metrics can be used to describe the \gls{ran} status. Since the number of downlink symbols and \glspl{tb}, or the \gls{mcs} and the buffer occupancy for the \gls{embb} slice are highly correlated, using them to represent the state of the network only increases the dimensionality of the state vector without introducing additional information. Conversely, the buffer occupancy and the number of \glspl{tb} enrich the representation with low redundancy. Therefore, the \gls{drl} agents for the xApps in this paper consider as input metrics the number of \glspl{tb}, the buffer occupancy (or the ratio of \gls{prb} granted and requested, which has a high correlation with the buffer status), and the downlink rate.

\subsection{Comparing Different \gls{drl}-based \gls{ran} Control Strategies}
\label{sec:models}

Once the input metrics have been selected, the next step in the design of \gls{ml} applications involves the selection of the proper modeling strategy~\cite{oran-ml}. In this paper, we consider
\gls{ml} models for sequential decision making, and thus focus on \gls{drl} algorithms.

\begin{figure}[t]
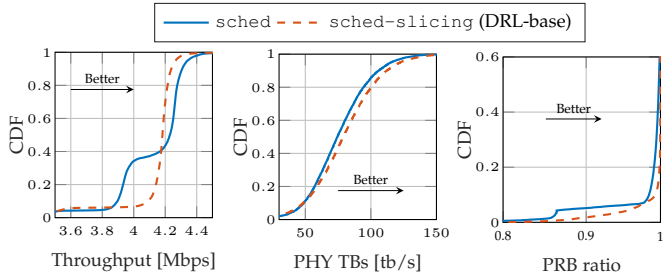

	\centering
	\begin{subfigure}[t]{0.33\columnwidth}
		\centering
		\setlength\fwidth{.75\columnwidth}
		\setlength\fheight{.75\columnwidth}
		\input{figures/sched-vs-sched-slicing-th-slice-0.tex}
		\caption{eMBB throughput.}
		\label{fig:th-sched-sched-slicing}
	\end{subfigure}\hfill
	\begin{subfigure}[t]{0.33\columnwidth}
		\centering
		\setlength\fwidth{.75\columnwidth}
		\setlength\fheight{.75\columnwidth}
		\input{figures/sched-vs-sched-slicing-pkt-slice-1.tex}
		\caption{MTC PHY TBs.}
		\label{fig:pkt-sched-sched-slicing}
	\end{subfigure}\hfill
	\begin{subfigure}[t]{0.33\columnwidth}
		\centering
		\setlength\fwidth{.75\columnwidth}
		\setlength\fheight{.75\columnwidth}
		\input{figures/sched-vs-sched-slicing-prb-slice-2.tex}
		\caption{URLLC PRB ratio.}
		\label{fig:prb-sched-sched-slicing}
	\end{subfigure}\hfill
	\caption{Comparison between the \texttt{sched} and \texttt{sched-slicing} xApps, with the slice-based traffic profile. The slicing for the \texttt{sched} xApp is fixed and based on the configuration chosen with highest probability by the \texttt{sched-slicing} xApp (36 \glspl{prb} for eMBB, 3 for MTC, 11 for URLLC).}
	\label{fig:sched-sched-slicing}
\end{figure}

\textbf{Control policy selection.} In this context, it is clearly crucial to properly select the control knobs, i.e., the \gls{ran} parameters that need to be controlled and adapted automatically, and the action space, i.e., the support on which these parameters can change. To this end, Fig.~\ref{fig:sched-sched-slicing} compares the performance for the \texttt{sched} and \texttt{sched-slicing} xApps, which perform different control actions. The first assumes a fixed slicing profile and includes three \gls{drl} agents that select the scheduling policy for each slice, while the second jointly controls the slicing (i.e., number of \glspl{prb} allocated to each slice) and scheduling policies with a single \gls{drl} agent. For this comparison, the slicing profile for the \texttt{sched} xApp evaluation matches the configuration that is chosen most often by the \texttt{sched-slicing} agent, and the source traffic is slice-based. The \glspl{cdf} of Fig.~\ref{fig:sched-sched-slicing} show that the joint control of slicing and scheduling improves the relevant metric for each slice, with the most significant improvements in the \gls{prb} ratio and in the
throughput for the users below the 40th percentile. This shows that there exist edge cases in which adapting the slicing profile further improves the network performance with respect to adaptive schedulers with a static slice configuration, even if the fixed slicing configuration is the one that is chosen most often by the \texttt{sched-slicing} xApp.

\begin{figure}[t]
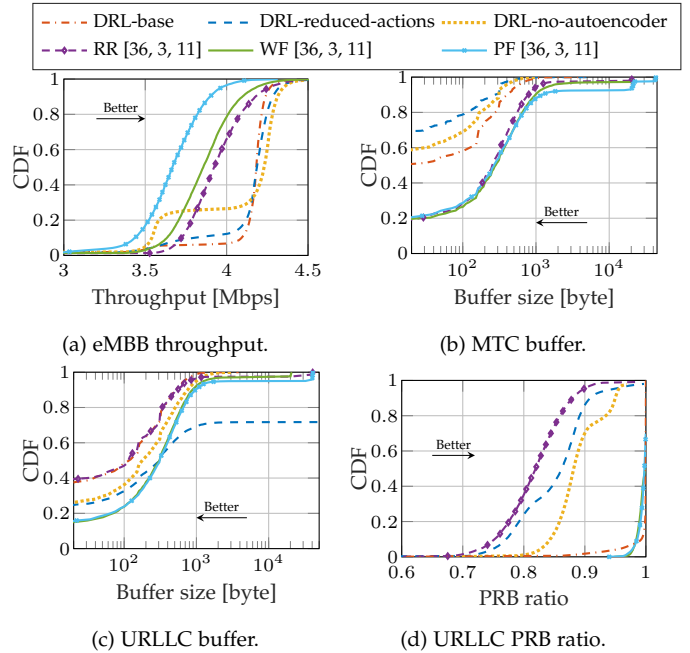

	\centering
	\begin{subfigure}[t]{0.48\columnwidth}
		\setlength\fwidth{.8\columnwidth}
		\setlength\fheight{.55\columnwidth}
		\input{figures/sched-slicing-model-comparison-th-slice-0.tex}
		\caption{eMBB throughput.}
		\label{fig:pkt-mtc}
	\end{subfigure}\hfill
	\begin{subfigure}[t]{0.48\columnwidth}
		\setlength\fwidth{.8\columnwidth}
		\setlength\fheight{.55\columnwidth}
		\input{figures/sched-slicing-model-comparison-buffer-slice-1.tex}
		\caption{\gls{mtc} buffer.}
		\label{fig:buf-mtc}
	\end{subfigure}
	\begin{subfigure}[t]{0.48\columnwidth}
		\setlength\fwidth{.8\columnwidth}
		\setlength\fheight{.55\columnwidth}
		\input{figures/sched-slicing-model-comparison-buffer-slice-2.tex}
		\caption{\gls{urllc} buffer.}
		\label{fig:buf-urllc}
	\end{subfigure}
	\begin{subfigure}[t]{0.48\columnwidth}
		\setlength\fwidth{.8\columnwidth}
		\setlength\fheight{.55\columnwidth}
		\input{figures/sched-slicing-model-comparison-prb-ratio-slice-2.tex}
		\caption{\gls{urllc} \gls{prb} ratio.}
		\label{fig:prb-urllc}
	\end{subfigure}
	\caption{Comparison between the different models of the \texttt{sched-slicing} xApp and baselines without \gls{drl}-based adaptation. For the latter, the performance is based on the slicing configuration chosen with highest probability by the best-performing \gls{drl} agent, and the three scheduler policies.}
	\label{fig:sched-slicing}
\end{figure}

\textbf{DRL agent design.} To further elaborate on the capabilities of
\texttt{sched-slicing},
in Fig.~\ref{fig:sched-slicing} we compare results for different configurations of the \gls{drl} agent of the xApp, as well as for a static baseline without
slicing or scheduling adaptation,
using the slice-based
traffic. 
The slicing profile for the static baseline
is the one
chosen most often by the \texttt{sched-slicing} xApp. The results of Fig.~\ref{fig:sched-slicing} further highlight the performance improvement introduced by adaptive, closed-loop control, with the \gls{drl}-driven control outperforming all baselines.

Additionally, this comparison spotlights the importance of careful selection of the action space for the \gls{drl} agents. By constraining or expanding the action space that the \gls{drl} agents can explore, the xApp designer can bias the selected policies. Consider the \gls{drl}-base and \gls{drl}-reduced-actions agents (see Table~\ref{tab:xapps}), whose difference is in the set of actions that the \gls{drl} agent can explore. Notably, the \gls{drl}-reduced-actions agent lacks the action that
results in the policy
chosen most often by the \gls{drl}-base agent. Compared to the most common action chosen by the \gls{drl}-reduced-actions agent (36 \gls{prb} for eMBB, 9 for MTC, 5 for URLLC), the most likely policy of \gls{drl}-base agent favors the URLLC over the MTC slice
(11 vs.\ 3 \glspl{prb}).
This is reflected in the performance metrics for the different slices. Notably, \gls{drl}-reduced-actions fails to maintain a small buffer and high \gls{prb} ratio for the \gls{urllc} slice (Fig.~\ref{fig:buf-urllc} and~\ref{fig:prb-urllc}), but achieves the smallest buffer occupancy for the \gls{mtc} traffic. 

\textbf{Autoencoder.} Finally, the results of Fig.~\ref{fig:sched-slicing} show the benefit of using an autoencoder, as the \gls{drl}-base and \gls{drl}-reduced-actions agents generally outperform the \gls{drl}-no-autoencoder agent. Indeed, the autoencoder decreases the dimensionality of the input for the \gls{drl} agent, improving the mapping between the network state and the actions. Specifically, the autoencoder used in this paper reduces a matrix of $T=10$ input vectors with $N=3$ metrics each to a single $N$-dimensional vector. Second, it improves the performance with online inference on real \gls{ran} data. Indeed, one of the issues
of operating \gls{ml} algorithms on live \gls{ran} telemetry is that some entries may be reported inconsistently or may be missing altogether. To address this, we train the autoencoder simulating the presence of a random number of zero entries in the training dataset. This allows the network to be able to meaningfully represent the state even
if
the input tensor is not fully populated with \gls{ran} data.

\section{Online Training for \gls{drl}-driven xApps}
\label{sec:online}

The last set of results presents an analysis of the tradeoffs associated with training \gls{drl} agents on a live network in an online fashion. These include the evaluation of the time required for convergence, the impact of the exploration process on the \gls{ran} performance, and the benefits involved with this procedure.
To do this, we load on the \texttt{online-training} xApp a model pre-trained on the offline dataset with the slice-based traffic profile. The same model is used in the \gls{drl}-reduced-actions agent. We deploy the \texttt{online-training} xApp on a \coloran base station and further continue the training with online exploration, using the uniform traffic profile (with the same constant bitrate traffic for each user).
Additionally, we leverage the containerized nature of \coloran to deploy it on Arena~\cite{bertizzolo2020arena}, a publicly available indoor testbed, and perform training with one \gls{sdr} base station and three
smartphones.

\begin{figure}[t]
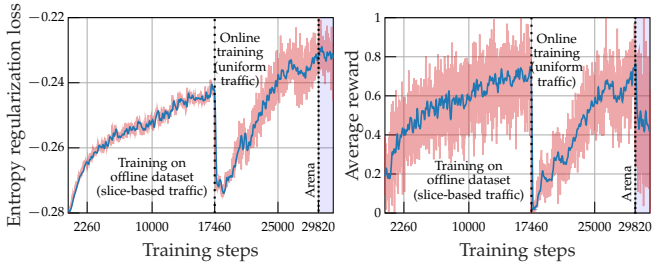

	\centering
	\begin{subfigure}[t]{0.495\columnwidth}
	\centering
    \setlength\abovecaptionskip{-.19cm}
	\setlength\fwidth{1.22\columnwidth}
	\setlength\fheight{.95\columnwidth}
	\input{figures/online-training-loss.tex}
	\caption{Entropy regularization loss.}
	\label{fig:loss}
	\end{subfigure}
	\hfill
	\begin{subfigure}[t]{0.495\columnwidth}
	\centering
	\setlength\fwidth{1.22\columnwidth}
	\setlength\fheight{.95\columnwidth}
	\input{figures/online-training-reward.tex}
	\caption{Reward.}
	\label{fig:reward}
	\end{subfigure}
	\caption{Metrics for the training on the offline dataset and the online training on Colosseum and Arena. The Arena configuration uses LTE band~7. Notice that the Arena deployment considers 3 users per base station, contrary to the 6 users per base station of Colosseum, thus the absolute average reward decreases.}
	\label{fig:reward-loss}
\end{figure}

\begin{figure}[t]
\begin{subfigure}[t]{0.7\columnwidth}
\setlength\fwidth{.8\textwidth}
\setlength\fheight{0.5\textwidth}
\includegraphics[width=\columnwidth]{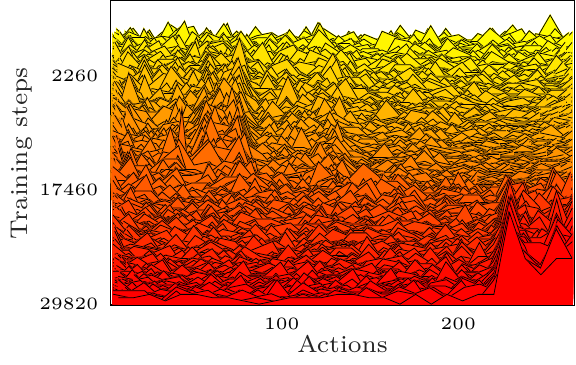}
\end{subfigure}
\begin{subfigure}[t]{0.2\columnwidth}
\setlength\fwidth{\textwidth}
\setlength\fheight{1.8\textwidth}
\includegraphics[width=1.13\columnwidth]{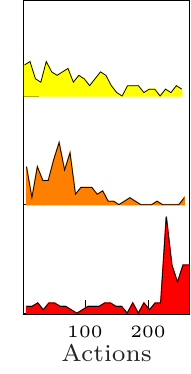}
\end{subfigure}
\caption{Distribution of the actions during the training on the offline dataset and the online training on Colosseum. The offline training stops at step 17460.}
\label{fig:histo}
\end{figure}

\textbf{Convergence.} Figures~\ref{fig:reward-loss} and~\ref{fig:histo} show how quickly the pre-trained agent adapts to the new environment. In particular, Fig.~\ref{fig:loss} reports the entropy regularization loss as a function of the training step of the agent. This metric correlates with the convergence of the training process: the smaller the absolute value of the entropy, the more likely the agent has converged
to
a set of actions that maximize the reward in the long run~\cite{haarnoja2017reinforcement}. We stop the training when this metric (and the average reward, Fig.~\ref{fig:reward}) plateaus, i.e., at step 17460 for the offline training, step 29820 for the online training on Colosseum. 
The loss remains stable when transitioning from the Colosseum to the Arena online training, while it increases (in absolute value) when switching traffic profile at step 17460. This shows that the agent can better generalize across different channel conditions than source traffic profiles. The same trend can be observed in the average reward (Fig.~\ref{fig:reward}), with the difference that the transition from Colosseum to Arena halves the reward (as this configuration features 3 and not 6 users for each base station). 
While the Colosseum online training requires 30\% fewer steps than
the initial offline training, it also comes with a higher wall-clock time. Indeed, offline exploration allows
the instantiation of
multiple parallel learning environments.
%
Because of this,
the Colosseum DGX
supports
the simultaneous exploration of 45 network configurations. Instead, online training can explore one configuration at a time, leading to a higher wall-clock time.

Figure~\ref{fig:histo} reports the evolution of the distribution of the actions chosen by the \gls{drl} agent for the Colosseum offline and online training. Three histograms for steps 2260, 17460 (end of offline training) and 29820 (end of online training) are also highlighted in the plot on the right. During training, the distribution of the actions evolves from uniform (in yellow) to more skewed, multi-modal distributions at the end of the offline training (in orange) and online training (in red). Additionally, when the training on the new environment begins, the absolute value of the entropy regularization loss increases (Fig.~\ref{fig:loss}), and, correspondingly, the distribution starts to change, until convergence to a new set of actions is reached again.

\begin{figure}[t]
	\centering
	\setlength\fwidth{.8\columnwidth}
	\setlength\fheight{.2\columnwidth}
	\input{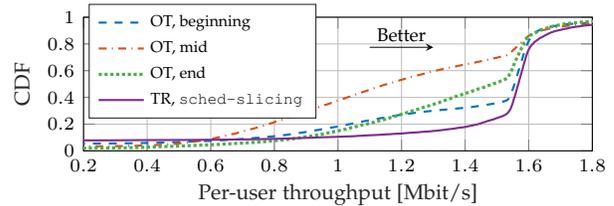}
	\caption{CDF of the throughput for the eMBB slice during the online training (OT) and with the trained agent (TR) with the uniform traffic profile.}
	\label{fig:cdf-online}
\end{figure}

\begin{figure}[t]
\centering
	\setlength\fwidth{.8\columnwidth}
	\setlength\fheight{.17\columnwidth}
	\input{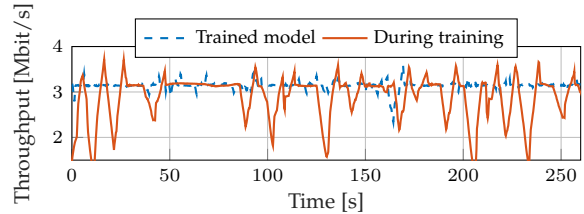}
	\caption{eMBB slice throughput during training and with the trained model.}
	\label{fig:th-online-online}
\end{figure}

\textbf{Impact of online training on RAN performance.} Achieving convergence with a limited number of steps is particularly important for online training, as the performance of the \gls{ran} may be negatively affected during the training process. Figure~\ref{fig:cdf-online} reports the \gls{cdf} for the user throughput during training and after, when the agent trained online is deployed on the \texttt{sched-slicing} xApp.
The performance worsens when comparing the initial training step, which corresponds to the agent still using the actions learned during offline training, with an intermediate step, in which
it is exploring random actions. Once the agent identifies the policies that maximize the reward in the new environment (in this case, with the uniform source traffic profile), the throughput improves. The best performance, however, is achieved with the trained agent, which does not perform any exploration. Figure~\ref{fig:th-online-online} further elaborates on this by showing how the online training process increases the throughput variability for the two eMBB users.
Therefore, performing online training on a production \gls{ran}~may be something a telecom operator cannot afford, as it may temporarily lead to disservices or reduced quality of service for the end users. In this sense, testbeds such as Colosseum can be an invaluable tool for two reasons. First, they provide the infrastructure to test pre-trained \gls{ml} algorithms---and \coloran enables any \gls{ran} developer to quickly onboard and test their xApps in a standardized O-RAN platform. Second, they allow online training
without affecting the performance of production environments.

\begin{figure}[t]
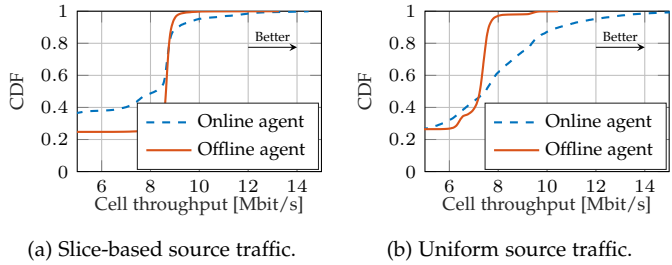

	\centering
	\begin{subfigure}[t]{0.48\columnwidth}
	    \centering
		\setlength\fwidth{.8\columnwidth}
		\setlength\fheight{.5\columnwidth}
		\input{figures/cdf-throughput-cell-traffic-75.tex}
		\caption{Slice-based source traffic.}
		\label{fig:cdf-th-75}
	\end{subfigure}\hfill
	\begin{subfigure}[t]{0.48\columnwidth}
	    \centering
		\setlength\fwidth{.8\columnwidth}
		\setlength\fheight{.5\columnwidth}
		\input{figures/cdf-throughput-cell-traffic-74.tex}
		\caption{Uniform source traffic.}
		\label{fig:cdf-th-74}
	\end{subfigure}

	\caption{Throughput comparison between the offline- and online-trained models with two source traffic patterns. The offline agent is the \gls{drl}-base for the \texttt{sched-slicing} xApp.}
	\label{fig:cdf-online-offline}
\end{figure}

\begin{figure}[t]
	\centering
		\setlength\fwidth{.75\columnwidth}
		\setlength\fheight{.2\columnwidth}
		\input{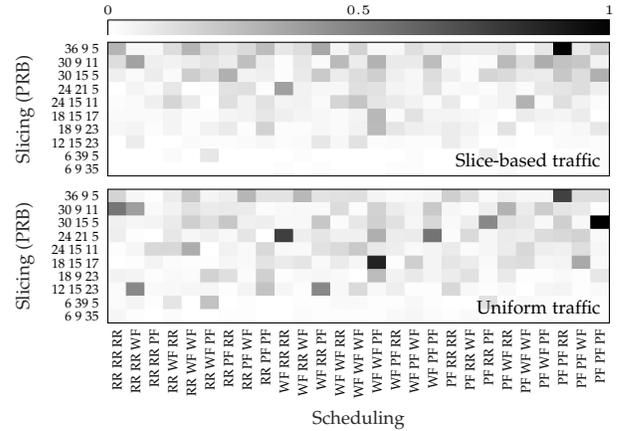}

		\caption{Probability of selecting a slicing/scheduling combination for the online-trained agent with two different source traffic patterns. For each tuple, the first element refers to the PRB (scheduling) for the eMBB slice, the second for the MTC slice, and the third for the URLLC slice.}
		\label{fig:actions-75}
\end{figure}

\textbf{Adaptability.} The main benefit of
an online training phase is to allow the pre-trained agent to adapt to updates in the environment that are not part of the training dataset. In this case, the agent trained by the \texttt{online-training} xApp adapts to a new configuration in the slice traffic, i.e., the uniform traffic profile. Figure~\ref{fig:cdf-online-offline} compares the cell throughout for the agent before/after the online training, with the slice-based (Fig.~\ref{fig:cdf-th-75}) and the uniform traffic (Fig.~\ref{fig:cdf-th-74}). Notably, the online agent achieves a throughput comparable with that of the agent trained on the offline dataset with slice-based traffic, showing that---despite the additional training steps---it is still capable of selecting proper actions for this traffic profile. This can also be seen in Fig.~\ref{fig:actions-75}, which shows that the action selected most often grants the most \glspl{prb} to the eMBB slice (whose users have a traffic one order of magnitude higher than MTC and URLLC).

The online agent, however, outperforms the offline-trained agent with the uniform traffic profile, with a gap of $2\:\mathrm{Mbit/s}$ in the 80th percentile, demonstrating the effectiveness of the online training to adapt to the updated traffic. The action profile also changes when comparing slice-based and uniform traffic, with a preference toward more balanced \gls{prb} allocations.

\textbf{Summary.} These results show how online training can help pre-trained models evolve and meet the demands of the specific environment in which they are deployed, at the cost, however, of reduced \gls{ran} performance during training. This makes the case for further research in this area, to develop, for example, smart scheduling algorithms that can alternate training and inference/control steps according to the needs of the network operator. Additionally, we showed that models pre-trained on Colosseum can be effective also in over-the-air deployments, making the case for \coloran as a platform to train and test O-RAN \gls{ml} solutions in a controlled environment.

\section{Related Work}
\label{sec:related}

The application of \gls{ml} to wireless networks has received considerable attention in recent years.
Existing works span the full protocol stack, with applications to channel modeling,
PHY and MAC layers, \gls{ml}-based routing and transport, and data-driven applications~\cite{oshea2016learning,abbasloo2021wanna,kibria2018big}. 

Several papers review the potential and challenges of \gls{ml} for wireless networks, discussing open issues and potential solutions. 
Kibria et al.\ highlight different areas in which \gls{ml} and big data analytics can be applied to wireless networks~\cite{kibria2018big}.
Sun et al.~\cite{sun2019application} and Gunduz et al.~\cite{gunduz2019machine} review the key learning techniques that researchers have applied to wireless, together with open issues. 
Similarly, Chen et al.\ focus on artificial neural network algorithms~\cite{chen2019artificial}. 
Other reviews can be found in~\cite{wang2020,jian2017machine}. 
While these papers present a clear overview of open problems associated with learning in wireless networks, and sometimes include some numerical evaluations~\cite{fu2018artificial,xiong2019deep}, they do not provide results based on an actual large-scale deployment, as this paper does, thus missing key insights on using real data, with imperfections, and on using closed-loop control on actual radios. 

When it comes to cellular networks, \gls{ml} has been applied throughout the \gls{3gpp} protocol stack. 
Perenda et al.\ automatically classify modulation and coding schemes~\cite{perenda2021learning}. 
Their approach is robust with respect to modulation parameters that are not part of the training set---a typical problem in wireless networks. 
Again, at the physical layer, Huang et al.\ investigate learning-based link adaptation schemes for the selection of the proper \gls{mcs} for \gls{embb} in case of preemptive puncturing for \gls{urllc}~\cite{huang2021deep}. 
Others apply \gls{ml} to~5G network management and \gls{kpm} prediction~\cite{polese2018machine,bega2019deepcog,wang2017spatio}.
These papers, however, do not close the loop through the experimental evaluation of the control action or classification accuracy on real testbeds and networks.
Chuai et al.\ describe a large-scale, experimental evaluation on a production network, but the evaluation is limited to a single performance metric~\cite{chuai2019collaborative}. 

\gls{drl} has recently entered the spotlight as a promising enabler of self-adaptive \gls{ran} control.
Nader et al.\ consider a multi-agent setup for centralized control in wireless networks, but not in the context of cellular networks~\cite{nader2021resource}.
Wang et al.\ use \gls{drl} to perform handover~\cite{wang2018handover}. 
Other papers analyze the theoretical performance of \gls{drl} agents for medium access~\cite{wang2018deep} and user association~\cite{zhao2019deep}. 
Mollahasani et al.\ evaluate
actor-critic learning for scheduling~\cite{mollahasani2021actor}, and Zhou et al.\ applies Q-learning to \gls{ran} slicing~\cite{zhou2021ran}.
Chinchali et al.\ apply \gls{drl} to user scheduling at the base station level~\cite{chinchali2018cellular}. 
Differently from these papers, we analyze the performance of \gls{drl} agents with a closed loop, implementing the control actions on a software-defined testbed with an O-RAN compliant infrastructure to provide insights on how \gls{drl} agents impact a realistic cellular network environment. Finally,~\cite{bonati2021intelligence,niknam2020intelligent} consider \gls{ml}/\gls{drl} applications in O-RAN, but provide a limited evaluation of the \gls{ran} performance without specific insights and results on using \gls{ml}.

\section{Conclusions}
\label{sec:conclusions}

The paper presents the first large-scale evaluation of \gls{ml}-driv\-en O-RAN xApps for managing and controlling a cellular network.
%
To this purpose, we introduce \coloran, the implementation of the O-RAN architecture in the Colosseum network emulator. 
\coloran features a \gls{ran} E2 termination, a near-real-time \gls{ric} with three different xApps, and a non-real-time \gls{ric} for data storage and \gls{ml} training. 
We pledge to publicly release \coloran to enable O-RAN-based experiments in Colosseum together with the dataset collected for this work. 
We demonstrate the effectiveness of \coloran through results from the large-scale comparative performance evaluation of the xApps running on \coloran and discuss key lessons learned on \gls{drl}-based closed-loop control.
In particular, we learned that (i) it is crucial to choose meaningful input features for the network state to avoid unnecessarily highly dimensional input for the \gls{drl} agent and that (ii) the action space for the \gls{drl} agent needs to be properly designed. 
Our comparison of different scheduling and slicing adaptation strategies shows that autoencoders can help to deal with unreliable real \gls{ran} data. 
Finally, we
provide insights on the live
training of \gls{drl} agents
in Colosseum and Arena.


\footnotesize  
\bibliographystyle{IEEEtran}
\bibliography{biblio.bib}

\begin{IEEEbiography}
[{\includegraphics[width=1in,height=1.25in,keepaspectratio]{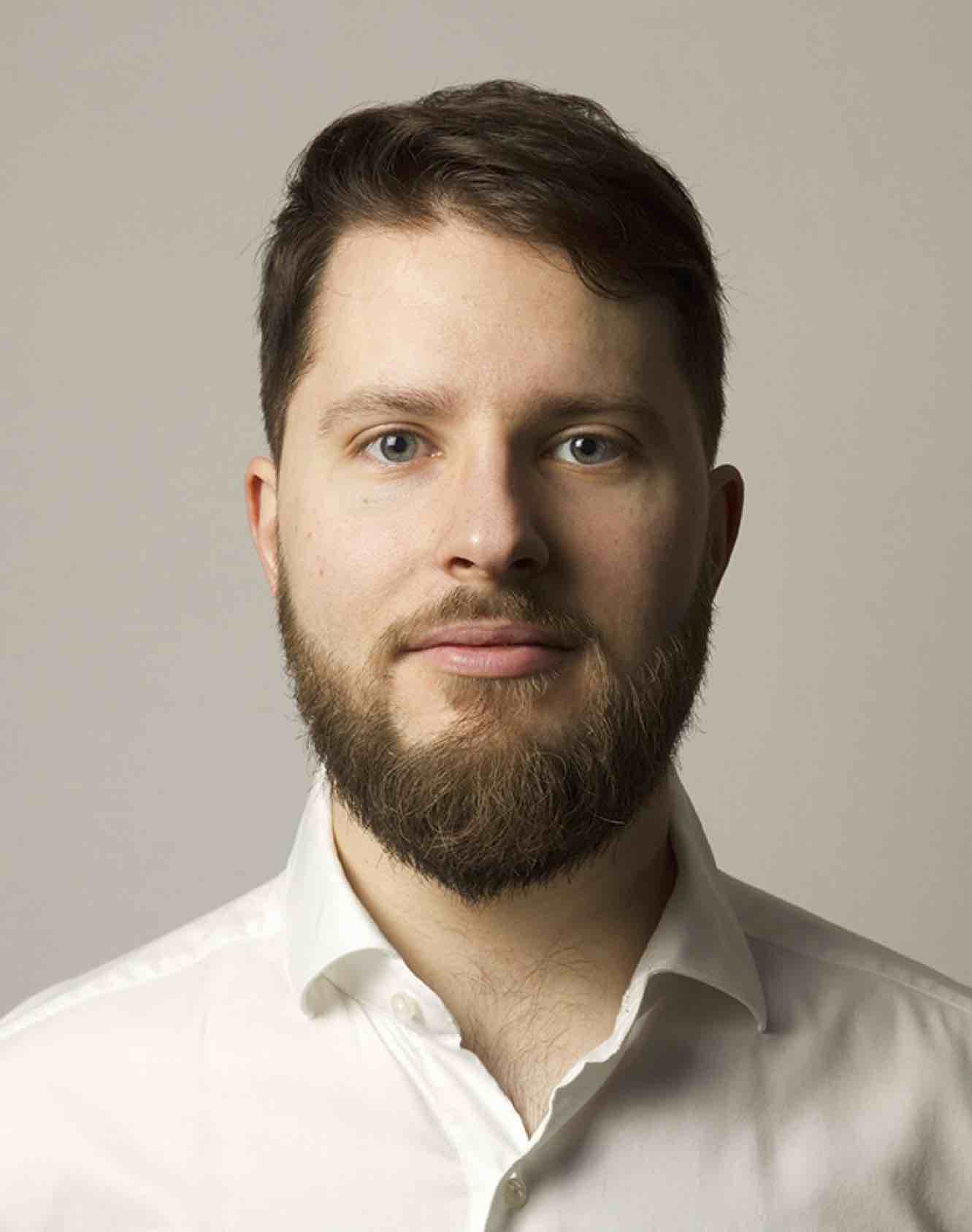}}]{Michele Polese} is an Associate Research Scientist at Northeastern University, Boston, since March 2020, working with Tommaso Melodia. He received his Ph.D. at the Department of Information Engineering of the University of Padova in 2020 under the supervision of with Michele Zorzi. He also was an adjunct professor and postdoctoral researcher in 2019/2020 at the University of Padova. During his Ph.D., he visited New York University (NYU), AT\&T Labs in Bedminster, NJ, and Northeastern University, Boston, MA. He collaborated with several academic and industrial research partners, including Intel, InterDigital, NYU, AT\&T Labs, University of Aalborg, King's College and NIST.
He was awarded with an Honorable Mention by the Human Inspired Technology Research Center (HIT) (2018), the Best Journal Paper Award of the IEEE ComSoc Technical Committee on Communications Systems Integration and Modeling (CSIM) 2019, and the Best Paper Award at WNS3 2019. His research interests are in the analysis and development of protocols and architectures for future generations of cellular networks (5G and beyond), in particular for millimeter-wave communication, and in the performance evaluation of complex networks. He is a Member of the IEEE.
\end{IEEEbiography}

\vspace{-.8cm}

\begin{IEEEbiography}
[{\includegraphics[width=1in,height=1.25in,keepaspectratio]{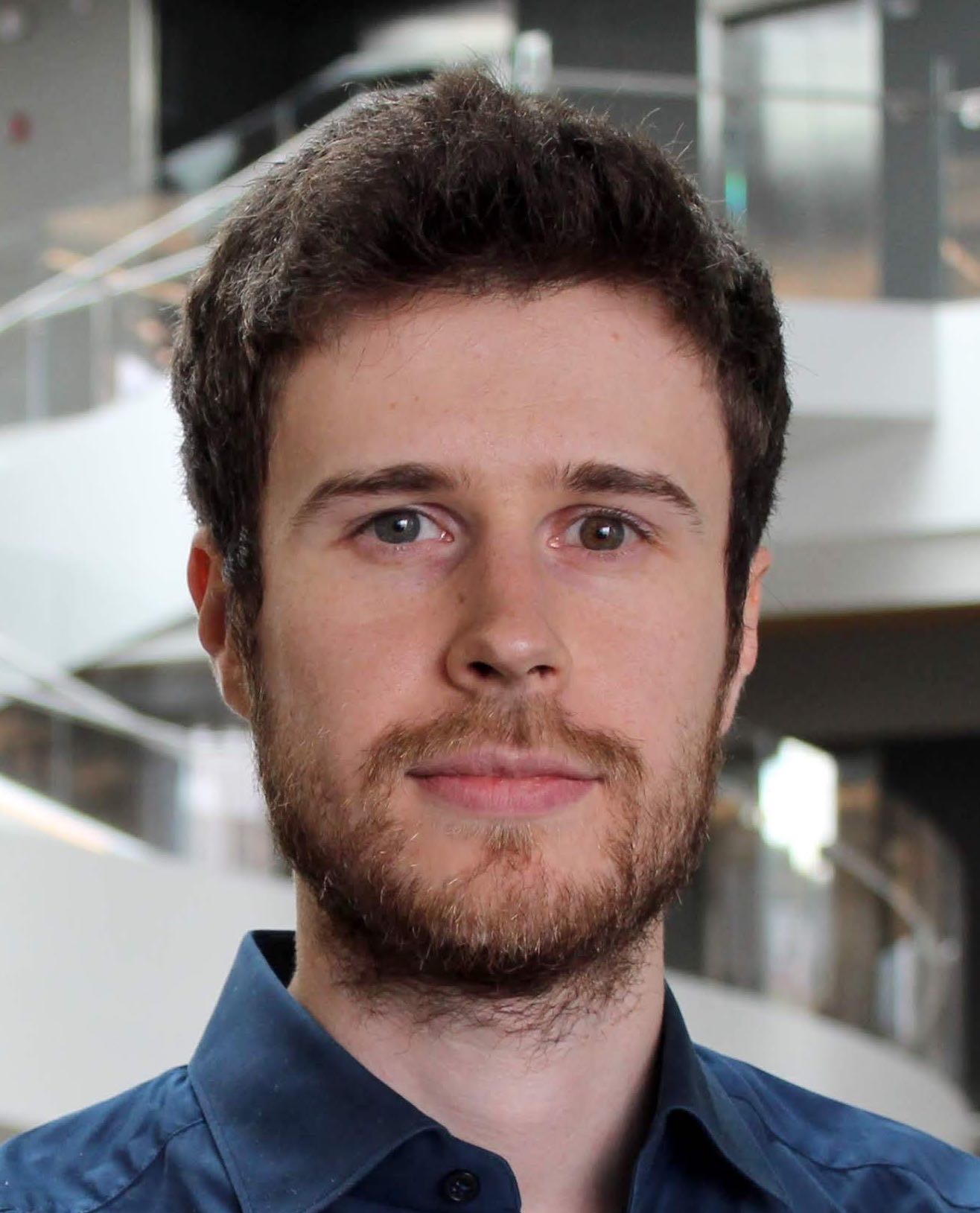}}]{Leonardo Bonati}
received his B.S. in Information Engineering and his M.S. in Telecommunication Engineering from University of Padova, Italy in 2014 and 2016, respectively. He is currently pursuing a Ph.D. degree in Computer Engineering at Northeastern University, MA, USA. His research interests focus on 5G and beyond cellular networks, network slicing, and software-defined networking for wireless networks.
\end{IEEEbiography}

\vspace{-.8cm}

\begin{IEEEbiography}
[{\includegraphics[width=1in,height=1.25in,keepaspectratio]{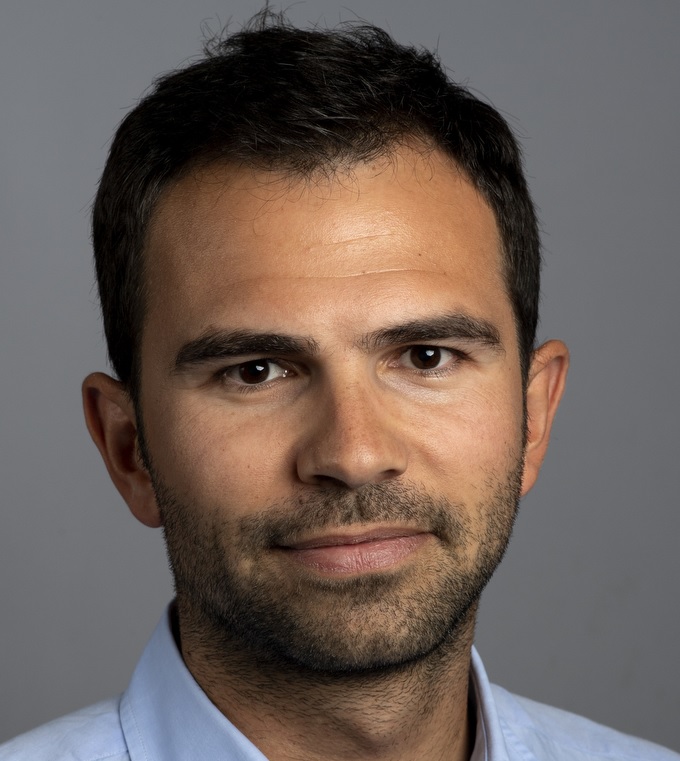}}]{Salvatore D'Oro}
received received his  Ph.D. degree from the University of Catania in 2015. He is currently a Research Assistant Professor at Northeastern University. He serves on the Technical Program Committee (TPC) of Elsevier Computer Communications journal and the IEEE Conference on Standards for Communications and Networking (CSCN) and European Wireless. He also served on the TPC of Med-Hoc-Net 2018 and several workshops in conjunction with IEEE INFOCOM and IEEE ICC. In 2015, 2016 and 2017 he organized the 1st, 2nd and 3rd Workshops on COmpetitive and COoperative Approaches for 5G networks (COCOA). 
Dr. D'Oro is also a reviewer for major IEEE and ACM journals and conferences. Dr. D'Oro's research interests include game-theory, optimization, learning and their applications to 5G networks. He is a Member of the IEEE.
\end{IEEEbiography}

\vspace{-.8cm}

\begin{IEEEbiography}
[{\includegraphics[width=1in,height=1.25in,keepaspectratio]{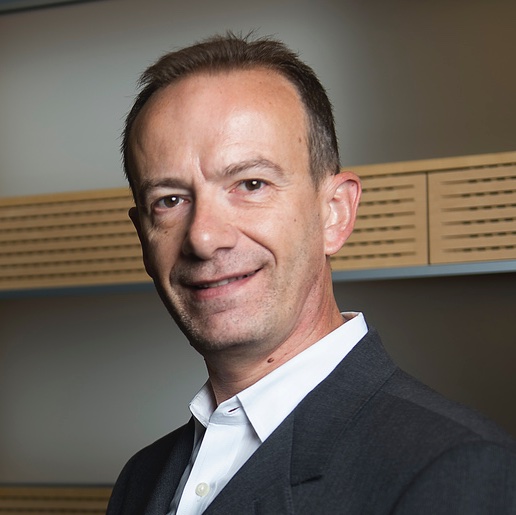}}]{Stefano Basagni}
is with the Institute for the Wireless Internet of Things and a professor at the ECE Department at Northeastern University, in Boston, MA. He holds a Ph.D.\ in electrical engineering from the University of Texas at Dallas (2001) and a Ph.D.\ in computer science from the University of Milano, Italy (1998). Dr. Basagni's current interests concern research and implementation aspects of mobile networks and wireless communications systems, wireless sensor networking for IoT (underwater, aerial and terrestrial), and definition and performance evaluation of network protocols.
Dr. Basagni has published over ten dozen of highly cited, refereed technical papers and book chapters. His h-index is currently 47 (December 2021). He is also co-editor of three books. Dr. Basagni served as a guest editor of multiple international ACM/IEEE, Wiley and Elsevier journals. He has been the TPC co-chair of international conferences. He is a distinguished scientist of the ACM, a senior member of the IEEE, and a member of CUR (Council for Undergraduate Education).
\end{IEEEbiography}

\begin{IEEEbiography}
[{\includegraphics[width=1in,height=1.25in,keepaspectratio]{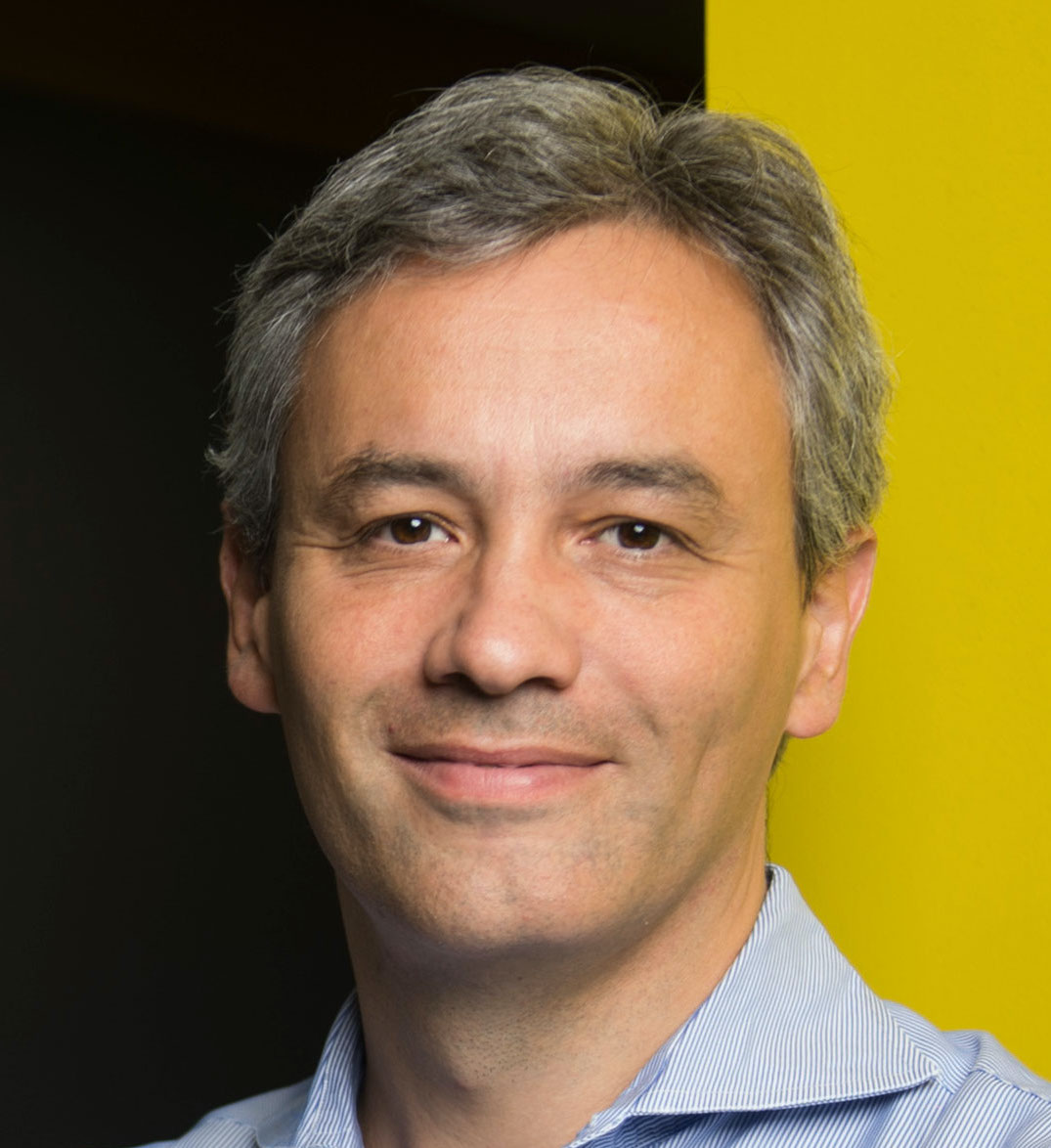}}]{Tommaso Melodia}
is the William Lincoln Smith Chair Professor with the Department of Electrical and Computer Engineering at Northeastern University in Boston. He is also the Founding Director of the Institute for the Wireless Internet of Things and the Director of Research for the PAWR Project Office. He received his Ph.D. in Electrical and Computer Engineering from the Georgia Institute of Technology in 2007. He is a recipient of the National Science Foundation CAREER award. Prof. Melodia has served as Associate Editor of IEEE Transactions on Wireless Communications, IEEE Transactions on Mobile Computing, Elsevier Computer Networks, among others. He has served as Technical Program Committee Chair for IEEE Infocom 2018, General Chair for IEEE SECON 2019, ACM Nanocom 2019, and ACM WUWnet 2014. Prof. Melodia is the Director of Research for the Platforms for Advanced Wireless Research (PAWR) Project Office, a \$100M public-private partnership to establish 4 city-scale platforms for wireless research to advance the US wireless ecosystem in years to come. Prof. Melodia's research on modeling, optimization, and experimental evaluation of Internet-of-Things and wireless networked systems has been funded by the National Science Foundation, the Air Force Research Laboratory the Office of Naval Research, DARPA, and the Army Research Laboratory. Prof. Melodia is a Fellow of the IEEE and a Senior Member of the ACM.
\end{IEEEbiography}

\vfill

\end{document}